\documentclass[aps,prd,amsfonts,amssymb, amsmath, showpacs, showkeys, a4paper, nofootinbib, superscriptaddress, 11pt, raggedbottom]{revtex4-2}

\usepackage{braket}
\usepackage{bm}
\usepackage{graphicx}
\usepackage{slashed}
\usepackage{subfig}
\usepackage{mathrsfs}
\usepackage{color}
\usepackage{mathtools}
\usepackage[normalem]{ulem}
\usepackage{simplewick}

\newcommand{\kket}[1]{| #1 \rangle\!\rangle}
\newcommand{\bbra}[1]{\langle\!\langle #1 |}
\newcommand{\bbrakket}[1]{\langle\!\langle #1 \rangle\!\rangle}

\def\beq{\begin{equation}}
\def\eeq{\end{equation}}
\def\beqa{\begin{eqnarray}}
\def\eeqa{\end{eqnarray}}

\begin{document}

\title{\Large A new method for directly computing reduced density matrices}

\author{Christian K\"{a}ding}
\email{christian.kaeding@tuwien.ac.at}
\affiliation{Technische Universit\"at Wien, Atominstitut, Stadionallee 2, 1020 Vienna, Austria}

\author{Mario Pitschmann}
\email{mario.pitschmann@tuwien.ac.at}
\affiliation{Technische Universit\"at Wien, Atominstitut, Stadionallee 2, 1020 Vienna, Austria}

\begin{abstract}
We demonstrate the power of a first principle-based and practicable method that allows for the perturbative computation of reduced density matrix elements of an open quantum system without making use of any master equations. The approach is based on techniques from non-equilibrium quantum field theory like thermo field dynamics, the Schwinger-Keldsyh formalism, and the Feynman-Vernon influence functional. It does not require the Markov approximation and is essentially a Lehmann-Szymanzik-Zimmermann-like reduction. In order to illustrate this method, we consider a real scalar field as an open quantum system interacting with an environment comprising another real scalar field. We give a general formula that allows for the perturbative computation of density matrix elements for any number of particles in a momentum basis. Finally, we consider a simple toy model and use this formula to obtain expressions for some of the system's reduced density matrix elements.

\end{abstract}


\maketitle

% =========================================================================================================================================

%\tableofcontents
%%%%%%%%%%%%%%%%%%%%%%%%%%%%%%%%%%

\section{Introduction}

Most realistic quantum systems can be regarded as open, which means that they are surrounded by a number of (uncontrollable) environments whose effects on the system cannot entirely be neglected \cite{Breuer2002}. Interactions between a system and its environments can lead to phenomena like momentum or energy diffusion \cite{Calzetta2008}, which may induce open quantum dynamical effects like phase shifts or decoherence \cite{Schlosshauer} if the system is in a superposed state. Applications of open quantum systems and their associated phenomena can be found in a variety of areas in non-relativistic quantum physics, see e.g.\,\,Ref.\,\cite{Carmichael,Gardiner2004,Walls2008,Aolita2015,Goold2016,Werner2016,Huber2020}, as well as in quantum field theory \cite{Calzetta2008,Koksma2010,Koksma2011,Sieberer2016,Marino2016,Baidya2017,Burrage2018,Nagy2020,Jana2021,Fogedby2022} and related areas, as, for example, Early Universe cosmology \cite{Lombardo1,Lombardo2,Lombardo3,Boyanovsky1,Boyanovsky2,Boyanovsky3,Boyanovsky4,Burgess2015,Hollowood,Binder2021}, black holes \cite{Yu2008,Lombardo2012,Jana2020,Agarwal2020,Kaplanek2020,Burgess2021,Kaplanek2021}, or heavy-ion physics \cite{Brambilla1,Brambilla2,Yao2018,Yao2020,Akamatsu2020,DeJong2020,Yao2021,Brambilla2021,Griend2021,Yao2022}. In addition, in recent years, there have been investigations of gravitationally-induced decoherence \cite{Blencowe,Anastopoulos2013,Oniga2015,Minar2016,Minar2016_2,Bassi2017,Asprea2019,Asprea2020,Asprea2021,Lagouvardos2021,Anastopoulos2021} and decoherence due to time dilation \cite{Pikovski2013,Pikovski2017}, as well as studies of the induction of open quantum dynamics due to interactions with environments comprising light scalar fields originating from modified gravity theories \cite{Burrage2018,Burrage2019,Kading2019}.
\\
In the theory of open quantum systems we usually deal with a system and its surrounding environment, which together form the total, closed system. As is the case in the present article, the open system could, for example, be a real quantum scalar field $\phi$ which is interacting with an environment in form of another real scalar field $\chi$. The total system consisting of $\phi$ and $\chi$ may be described via a density operator $\hat{\rho}_{\phi\chi}(t)$ fulfilling the quantum Liouville equation \cite{Breuer2002}. Tracing out the environmental degrees of freedom allows us to write down a quantum master equation for the elements of the reduced density matrix $\rho_\phi (t)$ in a basis of our choice (see Ref.\,\cite{Burrage2018} for a proper, first-principle way of doing this). While the evolution of the closed system is strictly unitary, $\rho_\phi (t)$ usually displays also non-unitary behavior leading to typical open quantum dynamical effects like phase shifts, or changes of coherence in form of re- or decoherence. Assuming a weak coupling between system and environment, and/or imposing the Markov approximation, which basically ignores memory effects in the environment due to its interaction with the system, might significantly simplify a master equation \cite{Breuer2002}.
\\
Nevertheless, actually solving a master equation often poses an intricate or even analytically impossible task. Therefore, especially if we are interested in making actual physical predictions for experiments, it might prove to be more useful to have a way of directly computing reduced density matrix elements without even having to write down a quantum master equation. That such a way actually exists, will be demonstrated by example in the present article. For this, we use the Lehmann-Szymanzik-Zimmermann- or LSZ-like reduction \cite{Lehmann1954} that was first presented in Ref.\,\cite{Burrage2018}, but also discussed in Refs.\,\cite{Burrage2019,Kading2019}. There, the authors used methods from non-equilibrium quantum field theory like thermo field dynamics (TFD) \cite{Takahasi:1974zn,Arimitsu:1985ez,Arimitsu:1985xm}, see also Ref.\,\cite{Khanna}, the Schwinger-Keldysh formalism \cite{Schwinger,Keldysh} and the Feynman-Vernon influence functional \cite{Feynman} in order to derive a master equation for a reduced density matrix in a single-particle momentum subspace. Using the well-known fact that the Schr\"odinger picture quantum Liouville equation may be expressed within TFD as a Schr\"odinger-like equation, which has well-understood solutions at the operator level, we apply the LSZ-like reduction for perturbatively finding explicit expressions for elements of the reduced density matrix $\rho_\phi$ in a momentum basis for any number of superposed or non-superposed particles in terms of the Feynman-Vernon influence functional. Clearly, for this, we must assume a weak coupling between system and environment. However, our approach allows for the description of non-Markovian open quantum systems.
\\
In principle, our method presented here also allows for the treatment of systems at finite temperatures by replacing the usual propagators by thermal ones (see e.g.\,\,Ref.\,\cite{LeBellac}), as was also done in Ref.\,\cite{Burrage2018}. Though, for simplicity and since we only want to demonstrate the general technique, we will not consider thermal effects in what follows. In addition, if we were particularly interested in discrete qudit systems, using the Schr\"odinger-like form of the quantum Liouville equation in TFD in combination with the results from Ref.\,\cite{Loubenets2020} might allow us to obtain exact, non-perturbative solutions in some situations.
\\
The present article is structured as follows: in Sec.\,\ref{sec:Derivation} we first introduce the necessary theoretical concepts and then use those in order to derive a general formula for the superposed single-particle element of the reduced density matrix $\rho_\phi(t)$ in a momentum basis. From this result we can then easily extrapolate a more general formula, which is valid for any number of particles. Next, in Sec.\,\ref{sec:Example}, we consider a simple, concrete example for the system field $\phi$ and the environmental field $\chi$. More specifically, we apply the derived formula in order to find explicit expressions for the associated vacuum and single-particle matrix elements of $\rho_\phi(t)$. Finally, in Sec.\,\ref{sec:Conclusion}, we draw our conclusions.

%%%%%%%%%%%%%%%%%%%%%%%%%%%%%%%%%%

\section{Derivation}
\label{sec:Derivation}

Here we are going to derive an expression that allows for the direct computation of reduced density matrix elements in terms of the Feynman-Vernon influence functional. Certainly, the methods presented in what follows can be applied to a variety of systems and environments. However, for simplicity and with applications to field theoretical systems in mind, we just consider a real scalar field $\phi$ as the system and another real scalar field $\chi$ as its surrounding environment. In order to better motivate the system-environment split, we may assume that $\phi$ has a constant mass $M$ much larger than $\chi$'s constant mass $m$, such that $m/M \ll 1$. This situation is, for example, comparable to the one described in Ref.\,\cite{Burrage2018}, where the system scalar field was used as a proxy for a heavy atom in atom interferometry that is interacting with an environment comprising fluctuations of a very light scalar field. 
\\
In Secs.\,\ref{ssec:Density}-\ref{ssec:TFD} we will first present the necessary mathematical and conceptual preliminaries before finally combining them for the actual derivation in Sec.\,\ref{ssec:EQN}.

%%%%%%%%%%%%%%%%%%%%%

\subsection{Density matrices in Fock space}
\label{ssec:Density}

Since we are working with a field theoretical system and want to be as general as possible, we must consider density matrix elements in Fock space, that allow for any number of particles. In order to closely follow the treatment in Ref.\,\cite{Burrage2018}, we choose to also work in a momentum basis and expand density operators accordingly. Consequently, the most general expression for a density operator expanded in this basis for any occupation number in Fock space is given by
\begin{eqnarray}\label{eq:DOFock}
\hat{\rho}(t) &=&  \sum\limits_{i,j=0}^\infty \frac{1}{i!j!} \int\left(   \prod\limits_{a = 1}^id\Pi_{\mathbf{k}^{(a)}}\right)\left(\prod\limits_{b = 1}^j   d\Pi_{\mathbf{l}^{(b)}} \right)\rho_{i;j}(\mathbf{k}^{(1)},...,\mathbf{k}^{(i)};\mathbf{l}^{(1)},...,\mathbf{l}^{(j)};t) \ket{\mathbf{k}^{(1)},...,\mathbf{k}^{(i)}}\bra{ \mathbf{l}^{(1)},...,\mathbf{l}^{(j)}}\,\,\,,\,\,\,\,\,\,\,\,\,\,\,
\end{eqnarray}
where the cases $i=0$ or $j=0$ correspond to the static vacuum state $\ket{0}$ or $\bra{0}$, respectively, each density matrix element is given by
\begin{eqnarray}
\rho_{i;j}(\mathbf{k}^{(1)},...,\mathbf{k}^{(i)};\mathbf{l}^{(1)},...,\mathbf{l}^{(j)};t) &=& \bra{\mathbf{k}^{(1)},...,\mathbf{k}^{(i)}}\hat{\rho}(t)\ket{ \mathbf{l}^{(1)},...,\mathbf{l}^{(j)}}\,\,\,,
\end{eqnarray}
and we use
\begin{eqnarray}
\int d\Pi_{\mathbf{k}} &:=& \int_{\mathbf{k}} \frac{1}{2E_{\mathbf{k}}}
\end{eqnarray}
with
\begin{eqnarray}
\int_{\mathbf{k}} &:=& \int \frac{d^3k}{(2\pi)^3}\,\,\,.
\end{eqnarray}
Later we will also make use of
\begin{eqnarray}
\int_{k} &:=& \int \frac{d^4k}{(2\pi)^4}\,\,\,.
\end{eqnarray}
We note that, as usual, the following has to hold:
\begin{eqnarray}\label{eq:PropDensMatr}
\rho_{i;j}(\mathbf{k}^{(1)},...,\mathbf{k}^{(i)};\mathbf{l}^{(1)},...,\mathbf{l}^{(j)};t) &=& \rho_{j;i}^\ast(\mathbf{l}^{(1)},...,\mathbf{l}^{(j)};\mathbf{k}^{(1)},...,\mathbf{k}^{(i)};t)\,\,\,.
\end{eqnarray}
Physically these density matrix elements can be interpreted as follows: $\rho_{0;0}$ describes a $0$-particle or vacuum state, $\rho_{i;0}$ or $\rho_{0;j}$ are correlations between the vacuum and $i$- or $j$-particle states, while $\rho_{i;j}$ represents correlations between $i$- and $j$-particle states.  
\\
As will become more apparent at a later point, it will be useful for us to initially describe the density operator in the Schr\"odinger picture. In this case, its time evolution is given by the quantum Liouville equation \cite{Breuer2002} (we use $\hbar \equiv 1$ throughout the entire article)
\begin{eqnarray}\label{Eqn:Liouville}
\frac{\partial}{\partial t} \hat{\rho}_S(t) &=& -\mathrm{i}[\hat{H}_S(t),\hat{\rho}_S(t)]\,\,\,,
\end{eqnarray}
which is solved by
\begin{eqnarray}\label{eqn:SPS1}
\hat{\rho}_S(t) &=& e^{-\mathrm{i}\hat{H}_S t}\hat{\rho}(0)e^{\mathrm{i}\hat{H}_S t}\,\,\,
\end{eqnarray}
if the Hamiltonian is time-independent, or by
\begin{eqnarray}\label{eqn:SPS2}
\hat{\rho}_S(t) &=& (\mathrm{T} e^{-\mathrm{i}\int^t_0 d\tau \hat{H}_S(\tau)})\hat{\rho}(0)(\tilde{\mathrm{T}} e^{\mathrm{i}\int^t_0 d\tau \hat{H}_S(\tau)})\,\,\,
\end{eqnarray}
if the Hamiltonian is time-dependent due to an external source, where the index $S$ labels objects in the Schr\"odinger picture, and $\mathrm{T}$ and $\tilde{\mathrm{T}}$ stand for time-ordering and anti-time-ordering, respectively. Note that Eqn.\,(\ref{eqn:SPS2}) is more general than the solution in Eqn.\,(\ref{eqn:SPS1}) and recovers the latter if $\hat{H}_S(t)$ is constant. 

%%%%%%%%%%%%%%%%%%%%%

\subsection{The Feynman-Vernon influence functional}
\label{ssec:FVIF}

\begin{figure}[htbp]
\begin{center}
\includegraphics[scale=0.5]{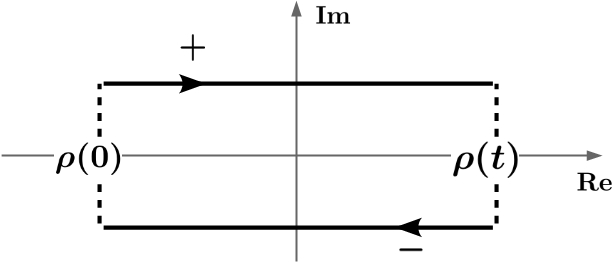}
\caption{Schematic depiction of the closed time path for a density matrix $\rho$ evolving from an initial time $0$ to a final time $t$ and backwards}
\label{fig:CTP}
\end{center}
\end{figure}

When describing an open system in field theory, the Feynman-Vernon influence functional \cite{Feynman} based on the Schwinger-Keldysh closed-time-path formalism \cite{Schwinger,Keldysh} is often the tool of choice, see e.g.\,\,Ref.\,\cite{Calzetta2008}. It is essentially relying on doubling the degrees of freedom, where the two copies are distinguished by labels $+$/$-$, and letting those evolve on the positive/negative branch of the closed time path depicted in Fig.\,\ref{fig:CTP} between an initial $t_\text{initial} =0$ and a final time $t_\text{final} =t$. Physically the Feynman-Vernon influence functional captures the evolution of the open system under influence of the environment, while not describing the environmental degrees of freedom directly. 
\\
Since the Feynman-Vernon influence functional will be an essential object in the equation that we are later going to derive, following the elaborations in and using the notations from Ref.\,\cite{Burrage2018}, we will now sketch how it comes into play when tracing out the environmental degrees of freedom of the total density operator. 
For this, we begin with the usual assumption that system and environment can be separated at the initial time, such that 
\begin{eqnarray}\label{eq:Initial}
\hat{\rho}_{\phi\chi}(0) &=& \hat{\rho}_\phi(0) \otimes \hat{\rho}_\chi(0)\,\,\,.
\end{eqnarray}
Since we are only interested in the evolution of the reduced density operator $\hat{\rho}_\phi(t)$, we trace out the environmental degrees of freedom:
\begin{eqnarray}
\hat{\rho}_{\phi}(t) &=& \mathrm{Tr}_\chi  \hat{\rho}_{\phi\chi}(t)\,\,\,.
\end{eqnarray}
Projecting this expression into a field basis living on the closed-time-path, we can express the reduced density functional as  
\begin{eqnarray}
\rho_\phi[\phi^\pm_t;t] &:=& \langle\phi^+_t|\hat{\rho}_\phi(t)|\phi^-_t\rangle
\nonumber
\\
&=& \int d\chi^\pm_t \delta(\chi^+_t - \chi^-_t) \rho_{\phi\chi}[\phi^\pm_t,\chi^\pm_t;t]\,\,\,,
\end{eqnarray}
where $\pm$ indicates a dependence on both $+$- and $-$-type operators, subscript $t$ labels the time slice on which the field eigenstates have to be taken, and 
\begin{eqnarray}
\rho_{\phi\chi}[\phi^\pm_t,\chi^\pm_t;t] &:=& 
\langle\phi^+_t,\chi^+_t|\hat{\rho}_{\phi\chi}(t)|\phi^-_t, \chi^-_t\rangle\,\,\,.
\end{eqnarray}
Furthermore, taking into account Eq.\,(\ref{eq:Initial}), the reduced density functional at time $t$ can be expressed as
\begin{eqnarray}
\rho_\phi[\phi^\pm_t;t] &=& \int d\phi_0^\pm \mathcal{I}[\phi^\pm_t,\phi^\pm_0;t,0] \rho_\phi[\phi^\pm_0;0]\,\,\,,
\end{eqnarray}
where the time translation is given in terms of the so-called influence functional (IF) propagator, see e.g.\,\,Ref.\cite{Calzetta2008},
\begin{eqnarray}\label{Eqn:IFProp}
\mathcal{I}[\phi^\pm_t,\phi^\pm_0;t,0] &=& \int^{\phi^\pm_t}_{\phi^\pm_0}\mathcal{D}\phi^\pm e^{\mathrm{i}\widehat{S}_\text{eff}[\phi;t]}\,\,\,.
\end{eqnarray}
The latter contains two path integrals over an effective action
\begin{eqnarray}\label{Eqn:EffS}
\widehat{S}_\text{eff}[\phi;t] &=& \widehat{S}_\phi[\phi;t] + \widehat{S}_{\phi,\text{int}}[\phi;t]+ \widehat{S}_\text{IF}[\phi;t]\,\,\,,
\end{eqnarray}
where ~$\widehat{}$~ indicates functionals that depend on both field variables $\phi^+$ and $\phi^-$. For example, the free action and the self-interaction of $\phi$ are given by
\begin{eqnarray}
\widehat{S}_\phi[\phi;t] &:=& S_\phi[\phi^+;t] - S_\phi[\phi^-;t]\,\,\,,
\nonumber
\\
\widehat{S}_{\phi,\text{int}}[\phi;t] &:=& S_{\phi,\text{int}}[\phi^+;t] - S_{\phi,\text{int}}[\phi^-;t]\,\,\,.
\end{eqnarray}
However, the influence action $\widehat{S}_\text{IF}[\phi;t]$ has a more involved definition by the Feynman-Vernon influence functional
\begin{eqnarray}\label{eq:InfluenceAction}
\widehat{\mathcal{F}}[\phi;t] &=& \exp\left\{ \mathrm{i}\widehat{S}_\text{IF}[\phi;t] \right\}
\nonumber
\\
&=& \int d\chi^\pm_t d\chi^\pm_0 \delta(\chi^+_t - \chi^-_t)\rho_\chi[\chi^\pm_0;0]
\nonumber
\\
&\phantom{=}&
\times \int^{\chi^\pm_t}_{\chi^\pm_0}\mathcal{D}\chi^\pm \exp\left\{ \mathrm{i}\Big( \widehat{S}_\chi[\chi;t] + \widehat{S}_{\chi,\text{int}}[\chi;t] + \widehat{S}_\text{int}[\phi,\chi;t] \Big) \right\}\,\,\,, 
\end{eqnarray}
where $\widehat{S}_\chi[\chi;t]$ is the free action, $\widehat{S}_{\chi,\text{int}}[\chi;t]$ the self-interaction action for $\chi$ and $\widehat{S}_\text{int}[\phi,\chi;t]$ an action describing the interaction between system and environment. Since we are working within a finite time interval, we define all actions used here only on $\Omega_t := [0,t]\times\mathbb{R}^3$, which means they all have the structure
\begin{eqnarray}
S[t]  &=& \int_{x\in\Omega_t} \mathcal{L}[x]
\end{eqnarray}
with 
\begin{eqnarray}
\int_x &:=& \int d^4x\,\,\,.
\end{eqnarray}
In addition, note that we are working with two types of functional integrals: one, denoted $d$, considers fields over all of $\mathbb{R}^3$, but only at one particular time slice, while the other, denoted $\mathcal{D}$, is over the whole $\Omega_t$. See Eq.\,(\ref{eq:InfluenceAction}) for an example in which both types appear together.
\\
Finally, we define an expectation value with respect to $\chi$ as (cf.\,\,Refs.\,\cite{Calzetta2008,Burrage2018})
\begin{eqnarray} 
\langle A[\chi^{a}]\rangle_\chi &:=& \int d\chi^{\pm}_t d\chi^{\pm}_0 \delta(\chi_t^+-\chi_t^-)\rho_\chi [\chi^{\pm}_0;0]
\int^{\chi^{\pm}_t}_{\chi^{\pm}_0} \mathcal{D}\chi^{\pm} A[\chi^{a}]\exp \left\{ \mathrm{i}\widehat{S}_{\chi}[\chi;t] \right\}\,\,\,,
\end{eqnarray} 
and find for the Feynman-Vernon influence functional:
\begin{eqnarray}\label{eqn:IFaction}
\widehat{\mathcal{F}}[\phi;t] &=& \left\langle \exp\big\{ \mathrm{i}\big( \widehat{S}_{\chi,\text{int}}[\chi;t] + \widehat{S}_\text{int}[\phi,\chi;t] \big) \big\} \right\rangle_\chi \,\,\,.
\end{eqnarray}

%%%%%%%%%%%%%%%%%%%%%

\subsection{Thermo field dynamics}
\label{ssec:TFD}

Next, we look at thermo field dynamics (TFD) \cite{Takahasi:1974zn,Arimitsu:1985ez,Arimitsu:1985xm} (see also Ref.\,\cite{Khanna}) as the final ingredient we need. Again we are following the description laid out in Ref.\,\cite{Burrage2018}. To some extent, we can understand TFD as an algebraic formulation of the Schwinger-Keldysh formalism, which itself works with operators on a positive $(+)$ or negative $(-)$ complex time axis, in a doubled Hilbert space
\begin{eqnarray}
\widehat{\mathcal{H}} &:=& \mathcal{H}^+ \otimes \mathcal{H}^-\,\,\,,
\end{eqnarray} 
where $\mathcal{H}^\pm$ are the Hilbert spaces corresponding to the $\pm$-branches of the closed time path, respectively.
Operators living on the closed time path can be expressed within TFD as
\begin{eqnarray}
\hat{\mathcal{O}}^+ &=& \hat{\mathcal{O}} \otimes \hat{\mathbb{I}}\,\,\,,
\nonumber
\\
\hat{\mathcal{O}}^- &=& \hat{\mathbb{I}} \otimes \hat{\mathcal{O}}^\mathcal{T}\,\,\,,
\end{eqnarray}
where $\hat{\mathbb{I}}$ is the unit operator and $\mathcal{T}$ means time reversal.
States in a momentum basis in TFD can be reached from the doubled vacuum state
\begin{eqnarray}
\kket{0} &:=& \ket{0} \otimes \ket{0}
\end{eqnarray}
via creation operators
\begin{eqnarray}\label{eq:Creators}
\hat{a}^{+\dagger}_\mathbf{k} \kket{0} \,=\, \ket{\mathbf{k}} \otimes \ket{0} \,=:\, \kket{\mathbf{k}_+}
\,\,\,,\,\,\,\,\,\,\,\,\,
\hat{a}^{-\dagger}_\mathbf{k} \kket{0} \,=\, \ket{0} \otimes \ket{\mathbf{k}} \,=:\, \kket{\mathbf{k}_-}\,\,\,.
\end{eqnarray}
Consequently, the corresponding annihilators act like
\begin{eqnarray}\label{eq:Annihilators}
\hat{a}^{\pm}_\mathbf{k} \kket{\mathbf{p}_+,\mathbf{p}_-} \,=\, (2\pi)^3 2 E^\phi_\mathbf{k}\delta^{(3)}(\mathbf{p}-\mathbf{k})\kket{\mathbf{p}_\mp}
\,\,\,.
\end{eqnarray}
In addition, we can construct a special state corresponding to the unit operator, see Ref.\,\cite{Arimitsu:1985ez},
\begin{eqnarray}\label{eq:SpecState}
\kket{1} &:=& \kket{0} + \int d\Pi_{\mathbf{k}} \kket{\mathbf{k}_+,\mathbf{k}_-} + \frac{1}{2!}\int d\Pi_{\mathbf{k}} d\Pi_{\mathbf{k'}}\kket{\mathbf{k}_+,\mathbf{k'}_+,\mathbf{k}_-,\mathbf{k'}_-}+...\,\,\,,
\end{eqnarray}
which allows us to express the expectation value of an operator as
\begin{eqnarray}
\braket{\hat{\mathcal{O}}(t)} &=& \text{Tr}\hat{\mathcal{O}}(t)\hat{\rho}(t) \,=\, \bbrakket{1| \hat{\mathcal{O}}^+(t)\hat{\rho}^+(t) |1}\,\,\,.
\end{eqnarray}
Having all this, we can now use the fact that Eq.\,(\ref{Eqn:Liouville}) can be re-written in a Schr\"odinger-like form:
\begin{eqnarray}\label{Eqn:S-likeLiou}
\frac{\partial}{\partial t} \hat{\rho}_S^+(t)\kket{1}_S &=& -\mathrm{i}\widehat{H}_S(t)\hat{\rho}_S^+(t)\kket{1}_S\,\,\,,
\end{eqnarray}
where $\widehat{H}_S(t) := \hat{H}_S(t) \otimes \hat{\mathbb{I}} - \hat{\mathbb{I}} \otimes \hat{H}_S(t)$. This can generally be solved by
\begin{eqnarray}\label{eq:SolutionTFD}
\hat{\rho}_S^+(t)\kket{1}_S &=& \text{T} \exp\left\{{-\mathrm{i}\int\limits_{0}^t\widehat{H}_S(\tau)d\tau}\right\}\hat{\rho}^+(0)\kket{1}_S\,\,\,.
\end{eqnarray}
Note that at time $0$ the different pictures coincide and we therefore dropped the label $S$.
\\
Analogously, we write down the Schr\"odinger-like form of the quantum Liouville equation for the reduced density operator that can be obtained by tracing out the $\chi$ degrees of freedom from the total density operator $\hat{\rho}_{\phi\chi}$:
\begin{eqnarray}\label{eq:QLiouRed}
\frac{\partial}{\partial t} \hat{\rho}_{\phi,S}^+(t)\kket{1}_S &=& -\mathrm{i}\widehat{H}_{\text{eff},S}(t)\hat{\rho}_{\phi,S}^+(t)\kket{1}_S\,\,\,,
\end{eqnarray}
where the effective Hamiltonian is given by the usual unitary evolution of $\phi$ due to the Hamiltonians $\hat{H}_{0,S}(t) + \hat{H}_{\text{int},S}(t)$ and a non-unitarian $\widehat{H}_{\text{IF},S}$:
\begin{eqnarray}
\widehat{H}_{\text{eff},S}(t) &=&  [\hat{H}_{0,S}(t) + \hat{H}_{\text{int},S}(t)] \otimes \hat{\mathbb{I}} - \hat{\mathbb{I}} \otimes [\hat{H}_{0,S}(t) + \hat{H}_{\text{int},S}(t)] + \widehat{H}_{\text{IF},S}(t)\,\,\,.
\end{eqnarray}
We recall that, generally, Hamiltonian and action are related via $H(t) = -\frac{\partial}{\partial t} S(t)$. Furthermore, if we consider that the effective action in Eq.\,(\ref{Eqn:EffS}) describes the full evolution of the reduced density matrix elements in the $\phi$-basis, the interaction picture $\widehat{H}_{\phi,I}$ must be its corresponding Hamiltonian, such that $\hat{H}_{0,I}(t) + \hat{H}_{\text{int},I}(t)$ represent $S_\phi(t) + S_{\phi,\text{int}}(t)$ and $\widehat{H}_{\text{IF},I}(t)$ the influence action given in Eq.\,(\ref{eq:InfluenceAction}). This will later be of greater importance when we write down a path integral expression for the reduced density matrix elements.
For now, however, we just remind ourselves that the free Hamiltonian $\hat{H}_0$ is the same in Schr\"odinger and interaction picture, and consequently drop the subscript.
\\
Since this will later be useful for us, we will now transform Eq.\,(\ref{eq:QLiouRed}) into the interaction picture. For this, we remind ourselves that an operator transforms like
\begin{eqnarray}
\hat{\mathcal{O}}_S &=& e^{-\mathrm{i}\hat{H}_0 t} \hat{\mathcal{O}}_I(t) e^{\mathrm{i}\hat{H}_0 t}
\end{eqnarray}
from the Schr\"odinger into the interaction picture, and conveniently define a unitary evolution operator 
\begin{eqnarray}
\widehat{U} &:=&  e^{-\mathrm{i}\hat{H}_0 t} \otimes e^{\mathrm{i}\hat{H}_0 t}
\,\,\,,
\end{eqnarray}
such that operators acting on the TFD doubled Hilbert space transform like
\begin{eqnarray}
\widehat{\mathcal{O}}_S &:=&  \widehat{U}\widehat{\mathcal{O}}_I(t)\widehat{U}^\dagger
\,\,\,.
\end{eqnarray}
The state introduced in Eq.\,(\ref{eq:SpecState}) transforms like a base vector:
\begin{eqnarray}
\widehat{U}^\dagger\kket{1}_S &=& \kket{1(t)}_I\,\,\,.
\end{eqnarray}
However, it can straightforwardly be seen that $\widehat{U}$ and $\widehat{U}^\dagger$ act like unit operators on $\kket{1}_S$ (and consequently on $\kket{1(t)}$ as well):
\begin{eqnarray}
\widehat{U}^\dagger\kket{1}_S &=& \ket{0}\otimes\ket{0} + \int d\Pi_{\mathbf{k}} e^{iE_\mathbf{k} t}\ket{\mathbf{k}}\otimes e^{-iE_\mathbf{k} t}\ket{\mathbf{k}}  + \frac{1}{2!}\int d\Pi_{\mathbf{k}} d\Pi_{\mathbf{k'}} e^{i(E_\mathbf{k}+E_\mathbf{k'}) t}\ket{\mathbf{k}}\otimes e^{-i(E_\mathbf{k}+E_\mathbf{k'}) t}\ket{\mathbf{k}}
+...
\nonumber
\\
&=& \kket{1}_S 
\,\,\,,
\end{eqnarray}
which implies that the state is time- and picture-independent, and will therefore simply be denoted by $\kket{1}$ in what follows. 
\\
Using all this, Eq.\,(\ref{eq:QLiouRed}) becomes in the interaction picture:
\begin{eqnarray}\label{eq:299}
\frac{\partial}{\partial t}\left( \widehat{U} \hat{\rho}_{\phi,I}^+(t) \widehat{U}^\dagger \right) \kket{1} &=& -\mathrm{i}\widehat{U}\widehat{H}_{\text{eff},I}(t)\hat{\rho}_{\phi,I}^+(t)\kket{1}\,\,\,.
\end{eqnarray}
Evaluating the partial derivative on the left-hand side of Eq.\,(\ref{eq:299}) leaves us with
\begin{eqnarray}
\left[(\partial_t \widehat{U}) \hat{\rho}_{\phi,I}^+(t) + \widehat{U}\partial_t\hat{\rho}_{\phi,I}^+(t) + \widehat{U}\hat{\rho}_{\phi,I}^+(t) \partial_t\widehat{U}^\dagger \right] \kket{1} &=& \widehat{U}\left[ -\mathrm{i}\widehat{H}_{0}\hat{\rho}_{\phi,I}^+(t) + \partial_t\hat{\rho}_{\phi,I}^+(t) + \mathrm{i}\hat{\rho}_{\phi,I}^+(t)\widehat{H}_{0} \right]\kket{1}
\,\,\,,\,\,\,\,\,\,\,\,\,\,\,\,\,\,\,
\end{eqnarray}
where
$\widehat{H}_{0} := \hat{H}_0\otimes \hat{\mathbb{I}} - \hat{\mathbb{I}}\otimes \hat{H}_0$. Acting with $\widehat{H}_{0}$ on $\kket{1}$ gives nil, as can straightforwardly be seen. Consequently, we find for Eq.\,(\ref{eq:299}):
\begin{eqnarray}
\left[ -\mathrm{i}\widehat{H}_{0}\hat{\rho}_{\phi,I}^+(t) + \partial_t\hat{\rho}_{\phi,I}^+(t) \right]\kket{1} &=& -\mathrm{i}\widehat{H}_{\text{eff},I}(t)\hat{\rho}_{\phi,I}^+(t)\kket{1}\,\,\,.
\end{eqnarray}
Since $\widehat{H}_{\text{eff},I}(t)$ also contains $\widehat{H}_{0}$, we finally arrive at
\begin{eqnarray}
 \partial_t\hat{\rho}_{\phi,I}^+(t)\kket{1} &=& -\mathrm{i}[\widehat{H}_{\text{int},I}(t)+\widehat{H}_{\text{IF},I}(t)]\hat{\rho}_{\phi,I}^+(t)\kket{1}\,\,\,,
\end{eqnarray}
where $\widehat{H}_{\text{int},I}(t) := \hat{H}_{\text{int},I}(t)\otimes \hat{\mathbb{I}} - \hat{\mathbb{I}} \otimes \hat{H}_{\text{int},I}(t)$,
and which can be solved by
\begin{eqnarray}\label{eq:AltSol}
\hat{\rho}_{\phi,I}^+(t)\kket{1}
&=& \text{T} \exp\left\{{-\mathrm{i}\int\limits_{0}^t[\widehat{H}_{\text{int},I}(\tau)+\widehat{H}_{\text{IF},I}(\tau)]d\tau}\right\}\hat{\rho}_{\phi}^+(0)\kket{1}
\,\,\,.
\end{eqnarray}

%%%%%%%%%%%%%%%%%%%%%

\subsection{Reduced density matrix elements}
\label{ssec:EQN}

Finally, we have all we need for deriving an equation that allows us to directly compute elements of the reduced density matrix $\rho_\phi(t)$ in a momentum basis for any occupation number in Fock space. For notational convenience, we will from now on drop the index $\phi$ when talking about the reduced density matrix and its elements. In addition, we will from now on only work in the interaction picture and therefore not use any corresponding labels for operators and states anymore.
\\
At first, we will derive an expression for the density matrix element $\rho_{1;1}(\mathbf{p};\mathbf{p}' ;t)$, which represents a single particle in momentum space. From the resulting equation and its derivation it will be straightforward for us to infer expressions for the other density matrix elements.
\\
As in Ref.\,\cite{Burrage2018}, we start with
\begin{eqnarray}\label{eq:335}
\langle\mathbf{p};t|\hat{\rho}(t)|\mathbf{p}';t\rangle &=& \rho_{1;1}(\mathbf{p};\mathbf{p}' ;t)\,\,\,.
\end{eqnarray}
Note that, as was pointed out in Refs.\,\cite{Millington:2012pf,Millington:2013isa}, the matrix element $\rho_{1;1}(\mathbf{p};\mathbf{p}\, ' ;t)$ is picture-independent.
In TFD, Eq.\,(\ref{eq:335}) can be expressed as 
\begin{eqnarray}
\text{Tr}|\mathbf{p}';t\rangle\langle\mathbf{p};t|\hat{\rho}(t) &=& \langle\langle 1|(|\mathbf{p}';t\rangle \langle\mathbf{p};t|\otimes \hat{\mathbb{I}})\hat{\rho}^+(t) |1\rangle\rangle\,\,\,.
\end{eqnarray}  
Substituting Eq.\,(\ref{eq:AltSol}), we are left with
\begin{eqnarray}\label{eq:zwiSchr1}
\rho_{1;1}(\mathbf{p};\mathbf{p}' ;t) &=&  \langle\langle \mathbf{p}_+,\mathbf{p}'_-;t| \text{T} \exp\left\{{-\mathrm{i}\int\limits_{0}^t[\widehat{H}_{\text{int}}(\tau)+\widehat{H}_{\text{IF}}(\tau)]d\tau}\right\}\hat{\rho}^+(0) |1\rangle\rangle\,\,\,.
\end{eqnarray}
Next, we expand the density operator in Eq.\,(\ref{eq:zwiSchr1}) as in Eq.\,(\ref{eq:DOFock}). However, we assume that the exact number of system particles and their correlations at the initial time are practically well-understood, such that we only have to consider the $\rho_{i;j}(0)$ elements for one particular choice of $i$ and $j$. In our case, we choose $i=j=1$ and assume all other initial elements to be nil. Though, if there was a reason to assume that the number of system particles is not certain, then the derivation could be easily modified. So, for the sake of readability, we do not consider the most general case here, but include it later when we extrapolate the most general expression. In our current case we find
\begin{eqnarray}
\rho_{1;1}(\mathbf{p};\mathbf{p} ' ;t) &=& \langle\langle \mathbf{p}_+,\mathbf{p}'_-;t| \text{T} \exp\left\{{-\mathrm{i}\int\limits_{0}^t[\widehat{H}_{\text{int}}(\tau)+\widehat{H}_{\text{IF}}(\tau)]d\tau}\right\}
\int d\Pi_{\mathbf{k}}d\Pi_{\mathbf{k}'}\rho_{1;1}(\mathbf{k};\mathbf{k}';0) |\mathbf{k}_+,\mathbf{k}'_-\rangle\rangle\,\,\,,\,\,\,\,\,\,\,\,\,\,\,\,\,\,\,
\end{eqnarray}
which, using Eqs.\,(\ref{eq:Creators}) and (\ref{eq:Annihilators}) becomes
\begin{eqnarray}
\rho_{1;1}(\mathbf{p};\mathbf{p}' ;t) &=& \int d\Pi_{\mathbf{k}}d\Pi_{\mathbf{k}'}\rho_{1;1}(\mathbf{k};\mathbf{k}';0)
\nonumber
\\
&&
\times
\langle\langle 0|\text{T}\hat{a}^+_{\mathbf{p}}(t) \hat{a}^-_{\mathbf{p} '}(t)
\exp\left\{{-\mathrm{i}\int\limits_{0}^t[\widehat{H}_{\text{int}}(\tau)+\widehat{H}_{\text{IF}}(\tau)]d\tau}\right\}
\hat{a}^{+\dagger}_{\mathbf{k}}(0) \hat{a}^{-\dagger}_{\mathbf{k} '}(0)|0\rangle\rangle\,\,\,.\,\,\,\,\,\,\,
\end{eqnarray}
After replacing the creation and annihilation operators by (cf.\,\,Ref.\,\cite{Burrage2018})
\begin{eqnarray}
\hat{a}^+_{\mathbf{p}}(t) &=& +\mathrm{i}\int_{\mathbf{x}} e^{-\mathrm{i}\mathbf{p}\cdot\mathbf{x}}\partial_{t,E^\phi_{\mathbf{p}}}\hat{\phi}^+(t,\mathbf{x})\,\,\,,\,\,\,\,\,\,\,\,\,
\hat{a}^{+\dagger}_{\mathbf{p}}(t) \,=\, -\mathrm{i}\int_{\mathbf{x}} e^{+\mathrm{i}\mathbf{p}\cdot\mathbf{x}} \partial_{t,E^\phi_{\mathbf{p}}}^*\hat{\phi}^+(t,\mathbf{x})\,\,\,,
\nonumber
\\
\hat{a}^-_{\mathbf{p}}(t) &=& -\mathrm{i}\int_{\mathbf{x}} e^{+\mathrm{i}\mathbf{p}\cdot\mathbf{x}}\partial_{t,E^\phi_{\mathbf{p}}}^*\hat{\phi}^-(t,\mathbf{x})\,\,\,,\,\,\,\,\,\,\,\,\,
\hat{a}^{-\dag}_{\mathbf{p}}(t) \,=\, +\mathrm{i}\int_{\mathbf{x}} e^{-\mathrm{i}\mathbf{p}\cdot\mathbf{x}}\partial_{t,E^\phi_{\mathbf{p}}}\hat{\phi}^{-}(t,\mathbf{x})\,\,\,,
\end{eqnarray}
where $\int_\mathbf{x}:=\int d^3x$ and $\partial_{t,E^\phi_{\mathbf{p}}} := \overset{\rightarrow}{\partial}_t - \mathrm{i}E^\phi_{\mathbf{p}}$,
we obtain
\begin{eqnarray}\label{eq:FinOpEx}
\rho_{1;1}(\mathbf{p};\mathbf{p}' ;t)
&=& \lim_{\substack{x^{0(\prime)}\,\to\, t^{+}\\y^{0(\prime)}\,\to\, 0^-}}
\int d\Pi_{\mathbf{k}} d\Pi_{\mathbf{k}'} \rho_{1;1}(\mathbf{k};\mathbf{k}';0) 
\nonumber
\\
&&
\times
\int_{\mathbf{x}\mathbf{x}'\mathbf{y}\mathbf{y}'}e^{-\mathrm{i}(\mathbf{p}\cdot\mathbf{x}-\mathbf{p}'\cdot\mathbf{x}')+\mathrm{i}(\mathbf{k}\cdot\mathbf{y}-\mathbf{k}'\cdot\mathbf{y}')}
\partial_{x^0,E^\phi_{\mathbf{p}}}\partial_{x^{0\prime},E^\phi_{\mathbf{p}'}}^*\partial_{y^0,E^\phi_{\mathbf{k}}}^*\partial_{y^{0\prime},E^\phi_{\mathbf{k}'}}
\nonumber
\\
&&
\times  
\langle\langle 0|{\rm T}[\hat{\phi}^+_x\hat{\phi}^-_{x'}\exp\left\{{-\mathrm{i}\int\limits_{0}^t[\widehat{H}_{\text{int}}(\tau)+\widehat{H}_{\text{IF}}(\tau)]d\tau}\right\} 
\hat{\phi}^{+}_y\hat{\phi}^{-}_{y'}]|0\rangle\rangle\,\,\,.\,\,\,\,\,
\end{eqnarray}
Here we introduced limits in which $x^{0(\prime)}$ approach $t$ from above and $y^{0(\prime)}$ approach $0$ from below in order to recover the correct time ordering, and started to use the short-hand notation $\phi_x := \phi(x)$ etc.
\\
Translating the resulting expression in Eq.\,(\ref{eq:FinOpEx}) into the path integral formalism, gives us
\begin{eqnarray}\label{eq:ZwiSchri2}
\rho_{1;1}(\mathbf{p};\mathbf{p}' ;t)  &=& 
\lim_{\substack{x^{0(\prime)}\,\to\, t^{+}\\y^{0(\prime)}\,\to\, 0^-}}
\int d\Pi_{\mathbf{k}} d\Pi_{\mathbf{k}'} \rho_{1;1}(\mathbf{k};\mathbf{k}';0) 
\nonumber
\\
&&
\times 
\int_{\mathbf{x}\mathbf{x}'\mathbf{y}\mathbf{y}'} e^{-\mathrm{i}(\mathbf{p}\cdot\mathbf{x}-\mathbf{p}' \cdot\mathbf{x}')+\mathrm{i}(\mathbf{k}\cdot\mathbf{y}-\mathbf{k}'\cdot\mathbf{y}')}
\partial_{x^0,E^\phi_{\mathbf{p}}} \partial_{x^{0\prime},E^\phi_{\mathbf{p}'}}^*\partial_{y^0,E^\phi_{\mathbf{k}}}^*\partial_{y^{0\prime},E^\phi_{\mathbf{k}'}}
\nonumber
\\
&&
\times 
\int\mathcal{D}\phi^{\pm} e^{\mathrm{i}\widehat{S}_{\phi}[\phi]}\phi^+_x\phi^-_{x'}
\exp\left\{{\mathrm{i}[\widehat{S}_{\phi,\text{int}}[\phi;t]+ \widehat{S}_\text{IF}[\phi;t]]}\right\}\phi^{+}_y\phi^{-}_{y'}\,\,\,,
\end{eqnarray}
where we used 
\begin{eqnarray}
\widehat{H}_{\text{int}}(\tau)+ \widehat{H}_{\text{IF}}(\tau) &=& \widehat{H}_\text{eff}(\tau) - \widehat{H}_0(\tau)
\nonumber
\\
&=&
\frac{\partial}{\partial \tau} \widehat{S}_\phi(\tau) - \frac{\partial}{\partial \tau} \widehat{S}_\text{eff}(\tau)
\nonumber
\\
&=&
\frac{\partial}{\partial \tau}\widehat{S}_\phi(\tau) + \sum_{a=\pm} a(\dot{\phi}^a \pi^a_\phi + \dot{\varphi}^a \pi^a_\varphi) - \left[\frac{\partial}{\partial \tau} \widehat{S}_\text{eff}(\tau) + \sum_{a=\pm} a(\dot{\phi}^a \pi^a_\phi + \dot{\varphi}^a \pi^a_\varphi)\right]
\nonumber
\\
&=&
\frac{d}{d \tau} \widehat{S}_\phi(\tau) - \frac{d}{d \tau} \widehat{S}_\text{eff}(\tau)
\nonumber
\\
&=&
-\frac{d}{d \tau}[\widehat{S}_{\phi,\text{int}}(\tau)+ \widehat{S}_\text{IF}(\tau)]
\,\,\,.
\end{eqnarray}

Using the definition in Eq.\,(\ref{eq:InfluenceAction}), we are finally led to
\begin{eqnarray}\label{eqn:11DensGen}
\rho_{1;1}(\mathbf{p};\mathbf{p}' ;t)
&=& 
\lim_{\substack{x^{0(\prime)}\,\to\, t^{+}\\y^{0(\prime)}\,\to\, 0^-}}
\int d\Pi_{\mathbf{k}} d\Pi_{\mathbf{k}'} \rho_{1;1}(\mathbf{k};\mathbf{k}';0) 
\nonumber
\\
&&
\times 
\int_{\mathbf{x}\mathbf{x}'\mathbf{y}\mathbf{y}'} e^{-\mathrm{i}(\mathbf{p}\cdot\mathbf{x}-\mathbf{p}' \cdot\mathbf{x}')+\mathrm{i}(\mathbf{k}\cdot\mathbf{y}-\mathbf{k}'\cdot\mathbf{y}')}
\partial_{x^0,E^\phi_{\mathbf{p}}} \partial_{x^{0\prime},E^\phi_{\mathbf{p}'}}^*\partial_{y^0,E^\phi_{\mathbf{k}}}^*\partial_{y^{0\prime},E^\phi_{\mathbf{k}'}}
\nonumber
\\
&&
\times 
\int\mathcal{D}\phi^{\pm} e^{\mathrm{i}\widehat{S}_{\phi}[\phi]}\phi^+_x\phi^-_{x'}\exp\left\{\mathrm{i}\widehat{S}_{\phi,\text{int}}[\phi;t] \right\}\widehat{\mathcal{F}}[\phi;t]\phi^{+}_y\phi^{-}_{y'}
\,\,\,.
\end{eqnarray}
What we just found is an equation that enables us to directly compute the momentum basis single-particle elements of the reduced density matrix $\rho_\phi(t)$ without having to explicitly solve any master equation. Using the assumption of a weak coupling would now allow us to perturbatively evaluate Eq.\,(\ref{eqn:11DensGen}) for a particular system-environment model in the same way as was done in Ref.\,\cite{Burrage2018} for a quantum master equation. Doing so will be subject of Sec.\,\ref{sec:Example}.
\\
Though, before concluding the present discussion, we will extrapolate the formula for a general density matrix element with an arbitrary choice of non-vanishing initial density matrix elements. That this result must be correct can easily be seen by retracing the steps we took in order to derive Eq.\,(\ref{eqn:11DensGen}). We find:
\begin{eqnarray}\label{eq:GenDensForm}
&&\rho_{g;h}(\mathbf{k}^{(1)},...,\mathbf{k}^{(g)};\mathbf{l}^{(1)},...,\mathbf{l}^{(h)};t)
= 
\nonumber
\\
&&
\sum\limits_{i,j=0}^\infty \frac{\mathrm{i}^{g+j}(-\mathrm{i})^{h+i}}{i!j!}
\lim_{\substack{x_{(1)}^{0},...,x_{(g)}^{0},x_{(1)}^{0\prime},...,x_{(h)}^{0\prime}\,\to\, t^{+}\\y_{(1)}^{0},...,y_{(i)}^{0},y_{(1)}^{0\prime},...,y_{(j)}^{0\prime}\,\to\, 0^-}}
\int \left(   \prod\limits_{a = 1}^id\Pi_{\mathbf{r}^{(a)}}\right)\left(\prod\limits_{b = 1}^j   d\Pi_{\mathbf{s}^{(b)}} \right) \rho_{i;j}(\mathbf{r}^{(1)},...,\mathbf{r}^{(i)};\mathbf{s}^{(1)},...,\mathbf{s}^{(j)};0) 
\nonumber
\\
&&
\times 
\int_{\mathbf{x}_{(1)}...\mathbf{x}_{(g)}\mathbf{x}_{(1)}^{\prime}...\mathbf{x}_{(h)}^{\prime}\mathbf{y}_{(1)}...\mathbf{y}_{(i)}\mathbf{y}_{(1)}^{\prime}...\mathbf{y}_{(j)}^{\prime}}
\exp\left\{-\mathrm{i}\Bigg(\sum\limits^g_{a=1}\mathbf{k}^{(a)}\mathbf{x}_{(a)}
-\sum\limits^h_{a=1}\mathbf{l}^{(a)}\mathbf{x}_{(a)}^{\prime}\Bigg)
+ \mathrm{i}\Bigg(\sum\limits^i_{a=1}\mathbf{r}^{(a)} \mathbf{y}_{(a)}-\sum\limits^j_{a=1}\mathbf{s}^{(a)} \mathbf{y}_{(a)}^{\prime}\Bigg)\right\}
\nonumber
\\
&&
\times 
\left(   \prod\limits_{a = 1}^g
\partial_{x^0_{(a)},E^\phi_{\mathbf{k}^{(a)}}} \right)
\left(  \prod\limits_{b = 1}^h
\partial_{x^{0\prime}_{(b)},E^\phi_{\mathbf{l}^{(b)}}}^*\right)
\left(\prod\limits_{c = 1}^i \partial_{y^0_{(c)},E^\phi_{\mathbf{r}^{(c)}}}^* \right)
\left(\prod\limits_{d = 1}^j 
\partial_{y^{0\prime}_{(d)},E^\phi_{\mathbf{s}^{(d)}}} 
\right)
\nonumber
\\
&&
\times 
\int\mathcal{D}\phi^{\pm} e^{\mathrm{i}\widehat{S}_{\phi}[\phi]}
\phi^+_{x_{(1)}}...\phi^+_{x_{(g)}}\phi^-_{x^\prime_{(1)}}...\phi^-_{x^\prime_{(h)}}
e^{\mathrm{i}\widehat{S}_{\phi,\text{int}}[\phi;t] }\widehat{\mathcal{F}}[\phi;t]
\phi^+_{y_{(1)}}...\phi^+_{y_{(i)}}\phi^-_{y^\prime_{(1)}}...\phi^-_{y^\prime_{(j)}}
\,\,\,.\,\,\,
\end{eqnarray}
Clearly, by taking $g=h=1$ and letting all intitial elements apart from the ones for $i=j=1$ vanish, we recover Eq.\,(\ref{eqn:11DensGen}).
\\
When evaluating Eqs.\,(\ref{eqn:11DensGen}) or (\ref{eq:GenDensForm}) it is important to understand that, at least for the system field $\phi$, only contractions of two $+$- or two $-$-fields are permitted, while for computing the influence functional via Eq.\,(\ref{eqn:IFaction}) all types of contractions of $\chi$ are allowed.
That the former is true for single-particle elements as in Eq.\,(\ref{eqn:11DensGen}) was already pointed out in Ref.\,\cite{Burrage2018}. There the authors showed that the $2\times 2$ matrix propagator used in their calculation is diagonal in the single-particle subspace at zero temperature. This is also the case in the calculation presented here. However, it can be seen that the matrix propagator is not only of diagonal form in the single-particle but also for all other $n$-particle subspaces. For this, we consider (schematically) the $g+h+i+j$-point function that was used in the derivation of Eq.\,(\ref{eq:GenDensForm}) as the generalisation of the four-point function in Eq.\,(\ref{eq:FinOpEx}) 
\begin{eqnarray}\label{eq:pointfunction}
\langle\langle 0|{\rm T}[\hat{\phi}^+_{x_{(1)}}...\hat{\phi}^+_{x_{(g)}}\hat{\phi}^-_{x^\prime_{(1)}}...\hat{\phi}^-_{x^\prime_{(h)}}\exp\left\{...\right\}
\hat{\phi}^+_{y_{(1)}}...\hat{\phi}^+_{y_{(i)}}\hat{\phi}^-_{y^\prime_{(1)}}...\hat{\phi}^-_{y^\prime_{(j)}}]|0\rangle\rangle
\,\,\,,
\end{eqnarray}
where $\exp\left\{...\right\}$ represents an exponentiation of a combination of $\phi$-field operators. After using Wick’s theorem \cite{Wick}, Eq.\,(\ref{eq:pointfunction}) reduces to a product of two-point functions. Taking into account the definition of the TFD vacuum state $\kket{0}$, it can be seen that the two-point functions that give the off-diagonal elements of the matrix propagator must vanish:
\begin{eqnarray}
\bbra{0}\hat{\phi}^{+(-)}_x\hat{\phi}^{-(+)}_y\kket{0}&=& 0\,\,\,.
\end{eqnarray}
The diagonal elements, however, give the usual Feynman and Dyson propagators
\begin{eqnarray}
\bbra{0}  \mathrm{T}[ \hat{\phi}^{+}_x\hat{\phi}^{+}_y ] \kket{0} &=& D^{++}_{xy} \,=\, D^\mathrm{F}_{xy} \,=\, - \mathrm{i}\int_k \frac{e^{\mathrm{i}k\cdot (x-y)}}{k^2+M^2-i\epsilon}\,\,\,,
\\
\bbra{0}  \mathrm{T}[ \hat{\phi}^{-}_x\hat{\phi}^{-}_y ] \kket{0} &=& D^{--}_{xy} \,=\, D^\mathrm{D}_{xy} \,=\, + \mathrm{i}\int_k \frac{e^{\mathrm{i}k\cdot (x-y)}}{k^2+M^2+i\epsilon}
\,\,\,.
\end{eqnarray}

%%%%%%%%%%%%%%%%%%%%%%%%%%%%%%%%%%

\section{A simple toy model}
\label{sec:Example}

Here we are going to apply the results from Sec.\,\ref{sec:Derivation}, i.e.\,\,Eqs.\,(\ref{eqn:11DensGen}) and (\ref{eq:GenDensForm}), to a toy model for system $\phi$ and environment $\chi$. For this, we choose a model that is as simple as possible, but still allows for a distinct system-environment split due to an enormous difference in both field's constant masses, and gives rise to at least some interesting open quantum dynamics.
\\
In Sec.\,\ref{ssec:Model} we will introduce the model and compute the resulting Feynman-Vernon influence functional perturbatively up to second order in the coupling constant. Next, in Sec.\,\ref{ssec:DMElements}, we will study a number of explicit cases: Firstly, we will look at the evolution of the vacuum matrix $\rho_{0;0}$, once with the initial condition $\rho_{0;0}(0) \neq 0$ and once with $\rho_{1;1}(0) \neq 0$. Secondly, we will investigate how the correlation between single particle and vacuum represented by $\rho_{1;0}$ evolves. For this, we assume that $\rho_{0;0}(0)$, $\rho_{1;0}(0)$, $\rho_{0;1}(0)$ and $\rho_{1;1}(0)$ are all non-vanishing. Finally, we will compute $\rho_{1;1}(t)$ under the initial condition $\rho_{1;1}(0) \neq 0$.
\\
After explicitly evaluating Eq.\,(\ref{eq:GenDensForm}) for each case, we will repeatedly encounter terms with time-dependent divergences. Dealing with such peculiar terms was briefly discussed in Ref.\,\cite{Burrage2018}. However, since we only intend to demonstrate how to use the formulas we derived in this article and not make actual experimental predictions for the model we use as an example, we will leave the divergent terms in the final results. An actual renormalization of such time-dependent divergences deserves a much more extensive discussion, which is beyond the scope of the present article, but will be the topic for a future one \cite{Kading2022}.

%%%%%%%%%%%%%%%%%%%%%

\subsection{The model}
\label{ssec:Model}

As an example we choose a two-scalar field model with no self-interactions, and the following actions:
\begin{eqnarray}
S_\phi[\phi] &=& \int_x \left[ -\frac{1}{2}(\partial\phi)^2 - \frac{1}{2}M^2\phi^2 \right]\,\,\,,
\\
S_\chi[\chi] &=& \int_x \left[ -\frac{1}{2}(\partial\chi)^2 - \frac{1}{2}m^2\chi^2 \right]\,\,\,,
\\
S_{\text{int}}[\phi,\chi] &=& \int_{x\in\Omega_t} \left[ -\frac{\alpha}{2}m \chi^2\phi \right]\,\,\,,
\end{eqnarray}
where $m/M, \alpha \ll 1$ and $\Omega_t := [0,t]\times\mathbb{R}^3$. Even though it is rather simple, this model already includes some interesting features that we would like to illustrate. Taking into account self-interactions and more coupling terms is therefore not much more illuminating, but more computationally challenging. A master equation for a density matrix element derived with the same LSZ-like reduction method used here but for a model with a self-interaction for $\chi$ and different coupling terms can be found in Ref.\,\cite{Burrage2018}.
\\
Using Eq.\,(\ref{eqn:IFaction}), we expand the Feynman-Vernon influence functional up to second order in $\alpha$:
\begin{eqnarray}\label{eq:FVIExpand}
\widehat{\mathcal{F}}[\phi;t] &=& \langle \exp\left\{ \mathrm{i} \widehat{S}_\text{int}[\phi,\chi;t]  \right\} \rangle_\chi
\nonumber
\\
&=& 1 + \mathrm{i}\sum\limits_{a=\pm} a\langle S_\text{int}[\phi^a,\chi^a;t] \rangle_\chi
- \frac{1}{2} \sum\limits_{a,b=\pm} ab \langle S_\text{int}[\phi^a,\chi^a;t] S_\text{int}[\phi^b,\chi^b;t]\rangle_\chi 
+ \mathcal{O}(\alpha^3)
\,\,\,.
\end{eqnarray}
In contrast to the system fields, where only Feynman and Dyson propagators are allowed, evaluating the expectation value with respect to $\chi$ includes all types of contractions and therefore also leads to the appearance of negative/positive frequency Wightman propagators. Note that here we consider zero temperature and the trivial $\chi$-vacuum, such that we have to use in what follows:
\begin{eqnarray}\label{eq:FeynmanProp}
\contraction{}{\chi}{^+_x}{\chi}\chi^+_x\chi^+_y
&=& \langle T\chi_{x} \chi_{y}\rangle  \,=\,  \Delta^{++}_{xy}  \,=\, \Delta^{\rm F}_{xy}
\,=\, - \mathrm{i} \int_k \frac{e^{ik\cdot(x-y)}}{k^2 + m^2 -\mathrm{i}\epsilon}\,\,\,,
\\
%%%%
\contraction{}{\chi}{^-_x}{\chi}\chi^-_x\chi^-_y
 &=& \langle\tilde{T} \chi_{x}\chi_{y}\rangle \,=\,  \Delta^{--}_{xy} \,=\, \Delta^{\rm D}_{xy}
\,=\, + \mathrm{i} \int_k  \frac{e^{ik\cdot(x-y)}}{k^2 + m^2 +\mathrm{i}\epsilon}\,\,\,,
\\
%%%%%
\contraction{}{\chi}{^+_x}{\chi}\chi^+_x\chi^-_y
&=& \langle\chi_{y} \chi_{x}\rangle \,=\,  \Delta^{+-}_{xy} \,=\, \Delta^<_{xy}
\,=\, \int_k e^{ik\cdot (x-y)} 2\pi \Theta(-k^0)\delta(k^2 + m^2)\,\,\,,
\\
%%%%
\label{eq:WMProp}
\contraction{}{\chi}{^-_x}{\chi}\chi^-_x\chi^+_y
&=&  \langle\chi_{x} \chi_{y}\rangle  \,=\, \Delta^{-+}_{xy} \,=\, \Delta^>_{xy}~=~ \Delta^<_{yx} \,=\, (\Delta^<)^*_{xy}\,\,\,.
\end{eqnarray}
These propagators fulfil the greatest time equation for any positive integer $n$:
\begin{eqnarray}
\sum_{a,b =\pm} ab (\Delta_{xy}^{ab})^n &=& 0\,\,\,.
\end{eqnarray}
Using Eqs.\,(\ref{eq:FeynmanProp})-(\ref{eq:WMProp}), we can separately compute each term in Eq.\,(\ref{eq:FVIExpand}) and find
\begin{eqnarray}
\label{eq:Sint}
\langle S_\text{int}[\phi^a,\chi^a;t] \rangle_\chi &=& -\frac{\alpha}{2}m\int_x \phi_x^a \Delta^{\rm F}_{xx}
\,\,\,,
\\
\label{eq:SintSint}
\langle S_\text{int}[\phi^a,\chi^a;t] S_\text{int}[\phi^b,\chi^b;t]\rangle_\chi
&=& \frac{\alpha^2}{4}m^2\int_{xy} \phi_x^a\phi_y^b\left[ \Delta^{\rm F}_{xx}\Delta^{\rm F}_{yy} + 2 (\Delta^{ab}_{xy})^2 \right]
\,\,\,.
\end{eqnarray}
Finally, substituting Eqs.\,(\ref{eq:Sint}) and (\ref{eq:SintSint}) into Eq.\,(\ref{eq:FVIExpand}), we obtain for the Feynman-Vernon influence functional:
\begin{eqnarray}\label{eq:FIVConcrete}
\widehat{\mathcal{F}}[\phi;t] 
&=& 1 -\mathrm{i}\frac{\alpha}{2}m\sum\limits_{a=\pm} a \int_x \phi_x^a \Delta^{\rm F}_{xx}
- \frac{\alpha^2}{8}m^2 \sum\limits_{a,b=\pm} ab \int_{xy} \phi_x^a\phi_y^b\left[ \Delta^{\rm F}_{xx}\Delta^{\rm F}_{yy} + 2 (\Delta^{ab}_{xy})^2 \right]
+ \mathcal{O}(\alpha^3)
\,\,\,.\,\,\,\,\,\,\,\,\,\,\,\,
\end{eqnarray}

%%%%%%%%%%%%%%%%%%%%%

\subsection{Density matrix elements}
\label{ssec:DMElements}

Now we are going to compute the elements of the density matrices $\rho_{0;0}$, $\rho_{1;0}$ and $\rho_{1;1}$ under different initial conditions. For this, we will use Eqs.\,(\ref{eq:GenDensForm}) and (\ref{eq:FIVConcrete}). How to evaluate these equations in detail was briefly outlined in Appendix D of Ref.\,\cite{Kading2019}.

%%%%%%%%%%%%%%%%%%%%%

\subsubsection{Vacuum state $\rho_{0;0}$}

\begin{figure}[htbp]
\centering
\subfloat[][]{\includegraphics[scale=0.35]{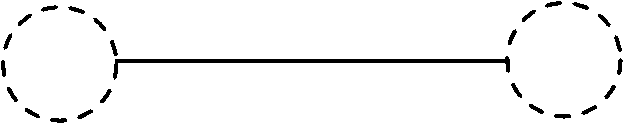}}
\qquad
\subfloat[][]{\includegraphics[scale=0.300]{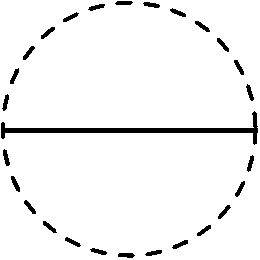}}
\caption{\label{Fig:Aleph} Diagrammatic representation of the terms contributing to $\aleph(t)$, as given in Eq.\,(\ref{eq:Aleph}); solid/dashed lines represent $\phi$/$\chi$-propagators. }
\end{figure}

We begin with the system-vacuum state under the assumption that only itself was non-vanishing at the initial time. In this case, Eq.\,(\ref{eq:GenDensForm}) becomes
\begin{eqnarray}
\rho_{0;0}(t)
&=& 
 \rho_{0;0}(0) 
\int\mathcal{D}\phi^{\pm} e^{\mathrm{i}\widehat{S}_{\phi}[\phi]}\widehat{\mathcal{F}}[\phi;t]\,\,\,.
\end{eqnarray}
Substituting Eq.\,(\ref{eq:FIVConcrete}) and contracting the system fields, leads us to 
\begin{eqnarray}\label{eq:InitVacuumState}
\rho_{0;0}(t)
&\approx&
\rho_{0;0}(0) \bigg\{ 1 
- \frac{\alpha^2}{8}m^2 \sum\limits_{a=\text{F},\text{D}}  \int_{zz'} \left[ \Delta^{a}_{zz}\Delta^{a}_{z'z'} + 2 (\Delta^{a}_{zz'})^2 \right]D^{a}_{zz'}
\bigg\}
\nonumber
\\
&=:&
\rho_{0;0}(0) [1+\aleph(t)]
\,\,\,,
\end{eqnarray}
where in the last line we summarized all terms beyond zeroth order as $\aleph(t)$.
As expected, the considered corrections to $\rho_{0;0}(0)$ are disconnected bubble diagrams, see Fig.\,\ref{Fig:Aleph}. It is obvious that this must also hold for all terms beyond second order in $\alpha$ since the initial vacuum density matrix in Eq.\,(\ref{eq:InitVacuumState}) can never connect to the Feynman-Vernon influence functional, leaving the latter's fields to contract only with themselves. Since we will also consider other cases, in which such disconnected diagrams will appear, we introduced the short-hand notation $\aleph(t)$, which can explicitly be evaluated as
\begin{eqnarray}\label{eq:Aleph}
\aleph(t) &\approx& 
-\frac{\alpha^2m^2}{2}
(2\pi)^3\delta^{(3)}(\mathbf{0}) \left\{
 \frac{\sin^2(Mt/2)}{M^3} \Delta^{\rm F}_{zz}\Delta^{\rm F}_{z'z'}
+\frac{1}{2}\int_{\mathbf{k}\mathbf{q}}  \frac{\sin^2[(E^\phi_\mathbf{k}+E^\chi_\mathbf{q}+E^\chi_{\mathbf{k}+\mathbf{q}})t/2]}{E^\phi_\mathbf{k}E^\chi_\mathbf{q}E^\chi_{\mathbf{k}+\mathbf{q}}(E^\phi_\mathbf{k}+E^\chi_\mathbf{q}+E^\chi_{\mathbf{k}+\mathbf{q}})^2}
\right\}
\,\,\,.\,\,\,\,\,\,\,\,\,\,\,\,
\end{eqnarray}
The first term on the right-hand side of Eq.\,(\ref{eq:Aleph}) corresponds to the tadpole diagram in Fig.\,\ref{Fig:Aleph}(a), while the second term corresponds to the vacuum bubble represented in Fig.\,\ref{Fig:Aleph}(b).

\begin{figure}[htbp]
\centering
\subfloat[][]{\includegraphics[scale=0.35]{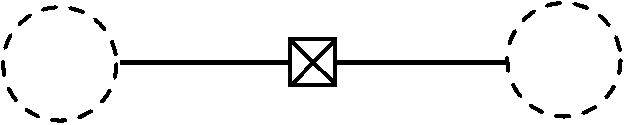}}
\qquad
\subfloat[][]{\includegraphics[scale=0.3]{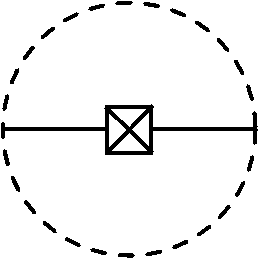}}
\caption{\label{Fig:rho0rho11} Diagrammatic representation of the terms contributing to $\rho_{0;0}(t)$ if only $\rho_{1;1}(0)$ was initially non-vanishing, as given in Eq.\,(\ref{eq:rho00rho11}); a crossed box depicts an insertion of an initial density matrix element, here of the matrix $\rho_{1;1}(0)$.}
\end{figure}

Next, we consider the case in which there existed a single particle at the initial time, represented by the matrix $\rho_{1;1}(0)$, with no vacuum correlations, and see how this modifies the vacuum density matrix at the final time $t$. For this, we write down the respective version of Eq.\,(\ref{eq:GenDensForm}) and substitute Eq.\,(\ref{eq:FIVConcrete}):
\begin{eqnarray}
\rho_{0;0}(t)
&=&
\lim_{y^{0(\prime)}\,\to\, 0^-}
\int d\Pi_{\mathbf{k}} d\Pi_{\mathbf{k}'} \rho_{1;1}(\mathbf{k};\mathbf{k}';0) 
\nonumber
\\
&&
\phantom{}
\times 
\int_{\mathbf{y}\mathbf{y}'} e^{\mathrm{i}(\mathbf{k}\cdot\mathbf{y}-\mathbf{k}'\cdot\mathbf{y}')}
\partial_{y^0,E^\phi_{\mathbf{k}}}^*\partial_{y^{0\prime},E^\phi_{\mathbf{k}'}}
\nonumber
\int\mathcal{D}\phi^{\pm} e^{\mathrm{i}\widehat{S}_{\phi}[\phi]}\widehat{\mathcal{F}}[\phi;t]\phi^{+}_y\phi^{-}_{y'}
\nonumber
\\
%%%%%%%
&\approx&
\frac{\alpha^2}{4}m^2
\lim_{y^{0(\prime)}\,\to\, 0^-}
\int d\Pi_{\mathbf{k}} d\Pi_{\mathbf{k}'} \rho_{1,1}(\mathbf{k},\mathbf{k}';0) 
\nonumber
\\
&&
\times 
\int_{\mathbf{y}\mathbf{y}'} e^{\mathrm{i}(\mathbf{k}\cdot\mathbf{y}-\mathbf{k}'\cdot\mathbf{y}')}
\partial_{y^0,E^\phi_{\mathbf{k}}}^*\partial_{y^{0\prime},E^\phi_{\mathbf{k}'}}
\int_{zz'} \big[ \Delta^\mathrm{F}_{zz}\Delta^\mathrm{F}_{z'z'} + 2 (\Delta^{+-}_{zz'})^2 \big]D^\mathrm{F}_{zy}D^\mathrm{D}_{z'y'}
\,\,\,.
%%%%%%%
\end{eqnarray}
In contrast to the previous case in Eq.\,(\ref{eq:InitVacuumState}), the fields coming from the Feynman-Vernon influence functional can now also contract with other fields, namely those coming from the initial density matrix. Therefore, we find:
\begin{eqnarray}\label{eq:rho00rho11}
\rho_{0;0}(t)
&\approx&
\frac{\alpha m^2}{4}\Bigg\{\frac{\sin^2(Mt/2)}{M^4}\Delta^\mathrm{F}_{zz}\Delta^\mathrm{F}_{z'z'}\rho_{1;1}(\mathbf{0};\mathbf{0};0)
\nonumber
\\
&&
\phantom{\approx\frac{\alpha m^2}{8}}
\left.
+\frac{1}{2}\int_{\mathbf{k}\mathbf{q}} \frac{\sin^2[(E^\phi_\mathbf{k}-E^\chi_\mathbf{q}-E^\chi_{\mathbf{k}-\mathbf{q}})t/2]}{(E^\phi_\mathbf{k})^2E^\chi_\mathbf{q}E^\chi_{\mathbf{k}-\mathbf{q}}(E^\phi_\mathbf{k}-E^\chi_\mathbf{q}-E^\chi_{\mathbf{k}-\mathbf{q}})^2}\rho_{1;1}(\mathbf{k};\mathbf{k};0)
\right\}\,\,\,.
\end{eqnarray}
These terms are diagrammatically represented in Fig.\,\ref{Fig:rho0rho11}. The first one on the right-hand side of Eq.\,(\ref{eq:rho00rho11}) corresponds to the two tadpoles in Fig.\,\ref{Fig:rho0rho11}(a) and the second one describes the bubble diagram given in Fig.\,\ref{Fig:rho0rho11}(b). We see that the right-hand side of Eq.\,(\ref{eq:rho00rho11}), including the time-dependent divergences, vanishes for $t\equiv0$, which is in concordance with the initial conditions.

%%%%%%%%%%%%%%%%%%%%%

\subsubsection{Correlation between single particle and vacuum $\rho_{1;0}$}

\begin{figure}[htbp]
\centering
\subfloat[][]{\includegraphics[scale=0.35]{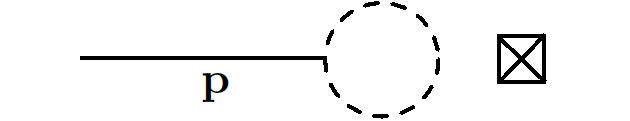}}
\qquad
\subfloat[][]{\includegraphics[scale=0.35]{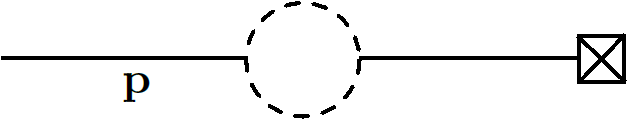}}

\subfloat[][]{\includegraphics[scale=0.35]{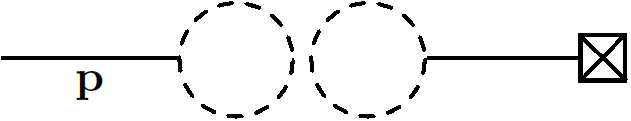}}
\qquad
\subfloat[][]{\includegraphics[scale=0.45]{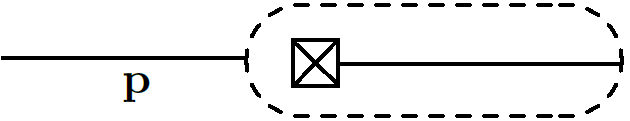}}

\subfloat[][]{\includegraphics[scale=0.45]{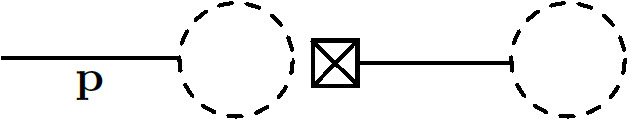}}
\qquad
\subfloat[][]{\includegraphics[scale=0.35]{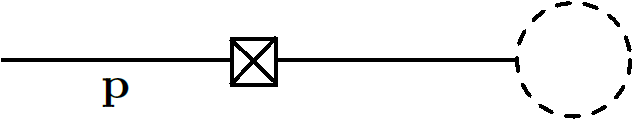}}
\caption{\label{Fig:rho10rho10} Diagrammatic representation of the terms contributing to $\rho_{1;0}(t)$ if $\rho_{0;0}(0)$ (Fig.\,(a)), $\rho_{1;0}(0)$ (Figs.\,(b) and (c)),  $\rho_{0;1}(0)$ (Figs.\,(d) and (e)), and $\rho_{1;1}(0)$ (Fig.\,(f)) were initially non-vanishing, as given in Eq.\,(\ref{eq:rho10rho10}) }
\end{figure}

We now move away from the vacuum state and have a look at the evolution of the correlation between a single particle and the vacuum. For this, we assume $\rho_{1;0}(0)$ to be non-vanishing. However, if such an initial correlation between a single particle and the vacuum existed, then we are required to also take into account non-vanishing $\rho_{0;0}(0)$ and $\rho_{1;1}(0)$. Furthermore, Eq.\,(\ref{eq:PropDensMatr}) tells us that also $\rho_{0;1}(0) \neq 0$. Consequently, we find for Eq.\,(\ref{eq:GenDensForm}) after substituting Eq.\,(\ref{eq:FIVConcrete}):
\begin{eqnarray}\label{eq:681}
\rho_{1;0}(\mathbf{p};t)
&=& 
\mathrm{i}\lim_{\substack{x^{0}\,\to\, t^{+}}} \rho_{0;0}(0) 
\int_{\mathbf{x}} e^{-\mathrm{i}\mathbf{p}\cdot\mathbf{x}}
\partial_{x^0,E^\phi_{\mathbf{p}}} 
\int\mathcal{D}\phi^{\pm} e^{\mathrm{i}\widehat{S}_{\phi}[\phi]}\phi^+_x\widehat{\mathcal{F}}[\phi;t]
\nonumber
\\
%%%%%%%
&&
+
\lim_{\substack{x^{0}\,\to\, t^{+}\\y^{0}\,\to\, 0^-}}
\int d\Pi_{\mathbf{k}}  \rho_{1;0}(\mathbf{k};0) 
\int_{\mathbf{x}\mathbf{y}} e^{-\mathrm{i}\mathbf{p}\cdot\mathbf{x}+\mathrm{i}\mathbf{k}\cdot\mathbf{y}}
\partial_{x^0,E^\phi_{\mathbf{p}}} \partial_{y^0,E^\phi_{\mathbf{k}}}^*
\int\mathcal{D}\phi^{\pm} e^{\mathrm{i}\widehat{S}_{\phi}[\phi]}\phi^+_x\widehat{\mathcal{F}}[\phi;t]\phi^{+}_y
\nonumber
\\
%%%%%%%
&&
-
\lim_{\substack{x^{0}\,\to\, t^{+}\\y^{0}\,\to\, 0^-}}
\int d\Pi_{\mathbf{k}}  \rho_{0;1}(\mathbf{k};0) 
\int_{\mathbf{x}\mathbf{y}} e^{-\mathrm{i}\mathbf{p}\cdot\mathbf{x}-\mathrm{i}\mathbf{k}\cdot\mathbf{y}}
\partial_{x^0,E^\phi_{\mathbf{p}}} \partial_{y^0,E^\phi_{\mathbf{k}}}
\int\mathcal{D}\phi^{\pm} e^{\mathrm{i}\widehat{S}_{\phi}[\phi]}\phi^+_x\widehat{\mathcal{F}}[\phi;t]\phi^{-}_y
\nonumber
\\
%%%%%%%
&&
+
\mathrm{i}\lim_{\substack{x^{0}\,\to\, t^{+}\\y^{0(\prime)}\,\to\, 0^-}}
\int d\Pi_{\mathbf{k}} \int d\Pi_{\mathbf{k}'}  \rho_{1;1}(\mathbf{k};\mathbf{k}';0) 
\int_{\mathbf{x}\mathbf{y}\mathbf{y}'} e^{-\mathrm{i}\mathbf{p}\cdot\mathbf{x}+\mathrm{i}(\mathbf{k}\cdot\mathbf{y}-\mathbf{k}'\cdot\mathbf{y}')}
\partial_{x^0,E^\phi_{\mathbf{p}}} \partial_{y^0,E^\phi_{\mathbf{k}}}^*\partial_{y^{0\prime},E^\phi_{\mathbf{k}'}}
\nonumber
\\
&&
\,\,\,\,\,\,
\times
\int\mathcal{D}\phi^{\pm} e^{\mathrm{i}\widehat{S}_{\phi}[\phi]}\phi^+_x\widehat{\mathcal{F}}[\phi;t]\phi^{+}_y\phi^{-}_{y'}
\,\,\,.
\end{eqnarray}
Next, we compute the path integrals and obtain
\begin{eqnarray}
\rho_{1;0}(\mathbf{p};t)
&\approx& 
\frac{\alpha}{2}m
\lim_{\substack{x^{0}\,\to\, t^{+}}} \rho_{0;0}(0) 
\int_{\mathbf{x}} e^{-\mathrm{i}\mathbf{p}\cdot\mathbf{x}}
\partial_{x^0,E^\phi_{\mathbf{p}}}
\int_z \Delta^\mathrm{F}_{zz} D^\mathrm{F}_{xz}
\nonumber
\\
%%%%%%%
&& 
+
\lim_{\substack{x^{0}\,\to\, t^{+}\\y^{0}\,\to\, 0^-}}
\int d\Pi_{\mathbf{k}}  \rho_{1;0}(\mathbf{k};0) 
\int_{\mathbf{x}\mathbf{y}} e^{-\mathrm{i}\mathbf{p}\cdot\mathbf{x}+\mathrm{i}\mathbf{k}\cdot\mathbf{y}}
\partial_{x^0,E^\phi_{\mathbf{p}}} \partial_{y^0,E^\phi_{\mathbf{k}}}^*
\nonumber
\\
&&
\,\,\,\,\,\,
\times
\bigg\{ [1+\aleph(t)]D^{\rm F}_{xy} 
-\frac{\alpha^2}{4}m^2 \int_{zz'}  \Big( \Delta^\mathrm{F}_{zz}\Delta^\mathrm{F}_{z'z'} + 2 (\Delta^\mathrm{F}_{zz'})^2 \Big)     D^\mathrm{F}_{xz}  D^\mathrm{F}_{z'y}
\bigg\} 
\nonumber
\\
%%%%%%%
&&
-
\frac{\alpha^2}{4}m^2
\lim_{\substack{x^{0}\,\to\, t^{+}\\y^{0}\,\to\, 0^-}}
\int d\Pi_{\mathbf{k}}  \rho_{0;1}(\mathbf{k};0) 
\int_{\mathbf{x}\mathbf{y}} e^{-\mathrm{i}\mathbf{p}\cdot\mathbf{x}-\mathrm{i}\mathbf{k}\cdot\mathbf{y}}
\partial_{x^0,E^\phi_{\mathbf{p}}} \partial_{y^0,E^\phi_{\mathbf{k}}}
\nonumber
\\
&&
\,\,\,\,\,\,
\times
\int_{zz'}  \Big( \Delta^\mathrm{F}_{zz}\Delta^\mathrm{D}_{z'z'} + 2 (\Delta^{+-}_{zz'})^2 \Big)     D^\mathrm{F}_{xz}  D^\mathrm{D}_{z'y}
\nonumber
\\
%%%%%%%
&& 
-\frac{\alpha}{2}m
\lim_{\substack{x^{0}\,\to\, t^{+}\\y^{0(\prime)}\,\to\, 0^-}}
\int d\Pi_{\mathbf{k}} \int d\Pi_{\mathbf{k}'}  \rho_{1;1}(\mathbf{k};\mathbf{k}';0) 
\nonumber
\\
&&
\,\,\,\,\,\,\,\,\,\,\,\,\,\,\,\,
\times 
\int_{\mathbf{x}\mathbf{y}\mathbf{y}'} e^{-\mathrm{i}\mathbf{p}\cdot\mathbf{x}+\mathrm{i}(\mathbf{k}\cdot\mathbf{y}-\mathbf{k}'\cdot\mathbf{y}')}
\partial_{x^0,E^\phi_{\mathbf{p}}} \partial_{y^0,E^\phi_{\mathbf{k}}}^*\partial_{y^{0\prime},E^\phi_{\mathbf{k}'}}
\int_z D^{\rm F}_{xy} D^{\rm D}_{zy'} \Delta^{\rm F}_{zz}
\,\,\,.
\end{eqnarray}
Notice how the term $\aleph(t)$, that we already computed in Eq.\,(\ref{eq:Aleph}), appears here as a dressing of the system propagator. Evaluating Eq.\,(\ref{eq:681}) leads us to  
\begin{eqnarray}
\label{eq:rho10rho10}
\rho_{1;0}(\mathbf{p};t)
&\approx&
-\frac{\alpha (2\pi)^3 }{2}\frac{m}{M}\delta^{(3)}(\mathbf{p}) \Delta^{\rm F}_{zz}\left( 1 - e^{-\mathrm{i}Mt} \right)  \rho_{0;0}(0)
%%%%
\nonumber
\\
&&
+
e^{-\mathrm{i}E^\phi_{\mathbf{p}} t}\rho_{1;0}(\mathbf{p};0)\Bigg\{
1 + \aleph(t) 
+\mathrm{i}\frac{\alpha^2m^2}{16E^\phi_{\mathbf{p}}}\sum\limits_{s=\pm}\int_\mathbf{q}\frac{1}{E^\chi_\mathbf{q} E^\chi_{\mathbf{p}-\mathbf{q}}(sE^\phi_{\mathbf{p}} + E^\chi_\mathbf{q} +E^\chi_{\mathbf{p}-\mathbf{q}})}
\nonumber
\\
&&
\,\,\,\,\,\,\,\,\,\,\,\,\,\,\,\,\,\,\,\,\,\,\,\,\,\,\,\,\,\,\,\,\,\,\,\,\,\,\,\,\,\,\,\,\,\,\,\,\,\,\,\,\,\,\,\,\,\,\,\,\,\,\,\,\,\,\,\,\,\,\,\,\,\,\,\,\,
\times
\Bigg[
t + \frac{\mathrm{i}}{sE^\phi_{\mathbf{p}} + E^\chi_\mathbf{q} +E^\chi_{\mathbf{p}-\mathbf{q}}}
\Bigg(
1 - e^{-\mathrm{i}(sE^\phi_{\mathbf{p}} + E^\chi_\mathbf{q} +E^\chi_{\mathbf{p}-\mathbf{q}})t}
\Bigg)
\Bigg]
\Bigg\}
\nonumber
\\
&&
+\frac{\alpha^2(2\pi)^3}{8}\frac{m^2}{M^3}\delta^{(3)}(\mathbf{p}) \Delta^\mathrm{F}_{zz}\Delta^\mathrm{F}_{z'z'}  \left( 1- e^{-\mathrm{i}Mt} \right)^2 \rho_{1;0}(\mathbf{0};0)
%%%%
\nonumber
\\
&&
-
\frac{\alpha^2m^2}{16E^\phi_{\mathbf{p}}}
e^{-\mathrm{i}E^\phi_{\mathbf{p}} t}\rho_{0;1}(-\mathbf{p};0)\int_\mathbf{q}\frac{\left(
1 - e^{\mathrm{i}(E^\phi_{\mathbf{p}} + E^\chi_\mathbf{q} +E^\chi_{\mathbf{p}-\mathbf{q}})t}
\right)\left(
1 - e^{\mathrm{i}(E^\phi_{\mathbf{p}} - E^\chi_\mathbf{q} -E^\chi_{\mathbf{p}-\mathbf{q}})t}
\right)}{E^\chi_\mathbf{q} E^\chi_{\mathbf{p}-\mathbf{q}}(E^\phi_{\mathbf{p}} + E^\chi_\mathbf{q} +E^\chi_{\mathbf{p}-\mathbf{q}})(E^\phi_{\mathbf{p}} - E^\chi_\mathbf{q} -E^\chi_{\mathbf{p}-\mathbf{q}})}
\nonumber
\\
&&
+\frac{\alpha^2(2\pi)^3}{2}\frac{m^2}{M^3}\delta^{(3)}(\mathbf{p}) \Delta^\mathrm{F}_{zz}\Delta^\mathrm{D}_{z'z'}  \sin^2(Mt/2) \rho_{0;1}(\mathbf{0};0)
%%%%
\nonumber
\\
&&
-\frac{\alpha m}{4M^2} \Delta^{\rm F}_{zz}\left( 1 - e^{\mathrm{i}Mt} \right) e^{-\mathrm{i}E^\phi_{\mathbf{p}} t} \rho_{1;1}(\mathbf{p};\mathbf{0};0)
\,\,\,.
\end{eqnarray}
In the first line of Eq.\,(\ref{eq:rho10rho10}) we see the creation of a single $\phi$-particle with momentum $\mathbf{p}$ out of the vacuum, see Fig.\,\ref{Fig:rho10rho10}(a). This process is possible since the $\chi$-degrees of freedom were traced out and are therefore not visible even though they are actually responsible for this creation via the process $\chi\chi \to \phi$. In the next two lines of Eq.\,(\ref{eq:rho10rho10}) we observe corrections to the unitary evolution of $\rho_{1;0}$ in form of disconnected diagrams corresponding to $\aleph(t)$, see Fig.\,\ref{Fig:Aleph}, and a loop correction to the propagator as depicted in Fig.\,\ref{Fig:rho10rho10}(b).
The first term in the fourth line of Eq.\,(\ref{eq:rho10rho10})  is only dependent on the initial $\rho_{1;0}$ for a particle at zero momentum in the chosen, fixed reference frame, and can be interpreted as a $\phi$-particle decaying within the time interval $[0,t)$ and another one with momentum $\mathbf{p}$ being produced before or at the final time $t$. This process is represented diagrammatically in Fig.\,\ref{Fig:rho10rho10}(c). The terms in the fifth and sixth lines of Eq.\,(\ref{eq:rho10rho10}) can be represented by the diagrams in Figs.\,\ref{Fig:rho10rho10}(d) and (e), which depict processes similar to those in Figs.\,\ref{Fig:rho10rho10}(b) and (c). Consequently, the terms in the fifth and sixth lines differ from those in the second, third and fourth lines even though there is some resemblance.  Finally, the last line of Eq.\,(\ref{eq:rho10rho10}) corresponds to the correlation between a unitarily evolving single particle and a particle decaying into the vacuum (via the process $\phi \to \chi\chi$), and is represented in Fig.\,\ref{Fig:rho10rho10}(f).

%%%%%%%%%%%%%%%%%%%%%

\subsubsection{Single-particle state $\rho_{1,1}$}
\begin{figure}[htbp]
\centering
\subfloat[][]{\includegraphics[scale=0.35]{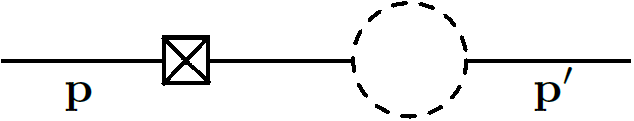}}
\qquad
\subfloat[][]{\includegraphics[scale=0.35]{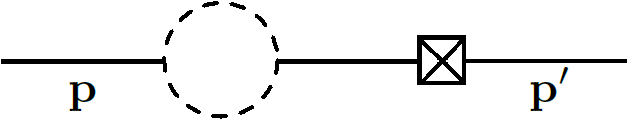}}

\subfloat[][]{\includegraphics[scale=0.35]{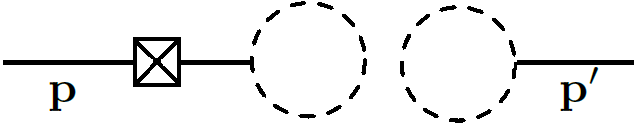}}
\qquad
\subfloat[][]{\includegraphics[scale=0.35]{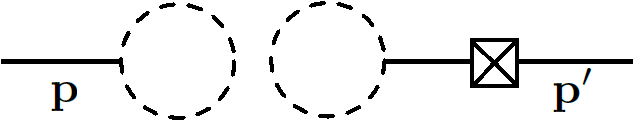}}

\caption{\label{Fig:rho11rho11} Diagrammatic representation of the terms contributing to $\rho_{1;1}(t)$ if only $\rho_{1;1}(0)$ was initially non-vanishing, as given in Eq.\,(\ref{eq:rho11rho11}) }
\end{figure}

Finally, we look at the evolution of a single particle in momentum space under the assumption that it already and only it existed at the initial time. In this case, we can use Eq.\,(\ref{eqn:11DensGen}), substitute Eq.\,(\ref{eq:FIVConcrete}), and find
\begin{eqnarray}
\rho_{1;1}(\mathbf{p};\mathbf{p}\, ' ;t)
&=&
\lim_{\substack{x^{0(\prime)}\,\to\, t^{+}\\y^{0(\prime)}\,\to\, 0^-}}
\int d\Pi_{\mathbf{k}} d\Pi_{\mathbf{k}'} \rho_{1;1}(\mathbf{k};\mathbf{k}';0) 
\nonumber
\\
&&
\times 
\int_{\mathbf{x}\mathbf{x}'\mathbf{y}\mathbf{y}\,'} e^{-\mathrm{i}(\mathbf{p}\cdot\mathbf{x}-\mathbf{p}\,' \cdot\mathbf{x}')+\mathrm{i}(\mathbf{k}\cdot\mathbf{y}-\mathbf{k}'\cdot\mathbf{y}\,')}
\partial_{x^0,E^\phi_{\mathbf{p}}} \partial_{x^{0\prime},E^\phi_{\mathbf{p}\,'}}^*\partial_{y^0,E^\phi_{\mathbf{k}}}^*\partial_{y^{0\prime},E^\phi_{\mathbf{k}'}}
\nonumber
\\
&&
\times 
\int\mathcal{D}\phi^{\pm} e^{\mathrm{i}\widehat{S}_{\phi}[\phi]}\phi^+_x\phi^-_{x'}\widehat{\mathcal{F}}[\phi;t]\phi^{+}_{y}\phi^{-}_{y'}
\,\,\,,
\end{eqnarray}
which becomes
\begin{eqnarray}\label{eq:788}
\rho_{1;1}(\mathbf{p};\mathbf{p}\, ' ;t)
&\approx&
\lim_{\substack{x^{0(\prime)}\,\to\, t^{+}\\y^{0(\prime)}\,\to\, 0^-}}
\int d\Pi_{\mathbf{k}} d\Pi_{\mathbf{k}'} \rho_{1;1}(\mathbf{k};\mathbf{k}';0) 
\nonumber
\\
&&
\times 
\int_{\mathbf{x}\mathbf{x}'\mathbf{y}\mathbf{y}'} e^{-\mathrm{i}(\mathbf{p}\cdot\mathbf{x}-\mathbf{p}' \cdot\mathbf{x}')+\mathrm{i}(\mathbf{k}\cdot\mathbf{y}-\mathbf{k}'\cdot\mathbf{y}')}
\partial_{x^0,E^\phi_{\mathbf{p}}} \partial_{x^{0\prime},E^\phi_{\mathbf{p}'}}^*\partial_{y^0,E^\phi_{\mathbf{k}}}^*\partial_{y^{0\prime},E^\phi_{\mathbf{k}'}}
\nonumber
\\
&&
\times
\bigg\{ [1+\aleph(t)] D^{\rm F}_{xy}D^{\rm D}_{x'y'} 
-\frac{\alpha^2}{4}m^2 \int_{zz'} \bigg[
\Big( \Delta^{\rm F}_{zz}\Delta^{\rm F}_{z'z'} + 2 (\Delta^{\rm D}_{zz'})^2 \Big) D^{\rm F}_{xy} D^{\rm D}_{x'z} D^{\rm D}_{z'y'}
\nonumber
\\
&&
+  \Big( \Delta^{\rm F}_{zz}\Delta^{\rm F}_{z'z'} + 2 (\Delta^{\rm F}_{zz'})^2 \Big) D^{\rm D}_{x'y'} D^{\rm F}_{xz'} D^{\rm F}_{zy}
\bigg]
\bigg\}\,\,\,.
\end{eqnarray}
Note the reappearance of $\aleph(t)$, which is now dressing the two $\phi$-propagators. Evaluating Eq.\,(\ref{eq:788}), we find: 
\begin{eqnarray}
\label{eq:rho11rho11}
\rho_{1;1}(\mathbf{p};\mathbf{p}' ;t)
&\approx&
e^{-\mathrm{i}(E^\phi_{\mathbf{p}}-E^\phi_{\mathbf{p}'}) t}\rho_{1;1}(\mathbf{p};\mathbf{p}';0)
\nonumber
\\
&&
\times
\Bigg\{1+\aleph(t)
+\mathrm{i}\frac{\alpha^2m^2}{16}\sum\limits_{s=\pm}\int_\mathbf{q}\frac{1}{E^\chi_\mathbf{q}}\Bigg[\frac{1}{E^\phi_{\mathbf{p}} E^\chi_{\mathbf{p}-\mathbf{q}}(sE^\phi_{\mathbf{p}} + E^\chi_\mathbf{q} +E^\chi_{\mathbf{p}-\mathbf{q}})}
\nonumber
\\
&&
\,\,\,\,\,\,\,\,\,\,\,\,\,\,\,\,\,\,\,\,\,\,\,\,\,\,\,\,\,\,\,\,\,\,\,\,\,\,\,\,\,\,\,\,\,\,\,\,\,\,\,\,\,\,\,\,\,\,\,\,\,\,\,\,\,\,\,\,\,\,\,\,\,\,\,\,\,\,
\times
\Bigg(
t + \frac{\mathrm{i}}{sE^\phi_{\mathbf{p}} + E^\chi_\mathbf{q} +E^\chi_{\mathbf{p}-\mathbf{q}}}
\Bigg(
1 - e^{-\mathrm{i}(sE^\phi_{\mathbf{p}} + E^\chi_\mathbf{q} +E^\chi_{\mathbf{p}-\mathbf{q}})t}
\Bigg)
\Bigg) 
\nonumber
\\
&&
\,\,\,\,\,\,\,\,\,\,\,\,\,\,\,\,\,\,\,\,\,\,\,\,\,\,\,\,\,\,\,\,\,\,\,\,\,\,\,\,\,\,\,\,\,\,\,\,\,\,\,\,\,\,\,\,\,\,\,\,\,\,\,\,\,\,\,\,\,\,\,\,\,\,\,\,\,\,
- (\mathbf{p}\longleftrightarrow\mathbf{p'})^\ast \Bigg]\Bigg\}
\nonumber
\\
&&
+ \frac{\alpha^2 m^2 (2\pi)^3 }{8M^3}\Delta^{\rm F}_{zz}\Delta^{\rm F}_{z'z'}\left[\delta^{(3)}(\mathbf{p}')\left( 1 - e^{\mathrm{i}Mt} \right)^2  e^{-\mathrm{i}E^\phi_{\mathbf{p}} t}\rho_{1;1}(\mathbf{p};\mathbf{0};0)
+ (\mathbf{p}\longleftrightarrow\mathbf{p'})^\ast
\right]
\,\,\,.
\end{eqnarray}
Again we find corrections to the unitary evolution, including de- and recoherence, in this case, of $\rho_{1;1}$. Those are given in the second to fourth lines of Eq.\,(\ref{eq:rho11rho11}) and comprise disconnected diagrams, as depicted in Fig.\,\ref{Fig:Aleph}, and loop corrections to the two propagators corresponding to the diagrams in Figs.\,\ref{Fig:rho11rho11}(a) and (b). In the last line of Eq.\,(\ref{eq:rho11rho11}) we only have terms proportional to initial density matrix elements with one of the two momenta vanishing in the chosen, fixed reference frame, and which correspond to the decay and creation of particles, as shown in Figs.\,\ref{Fig:rho11rho11}(c) and (d). The processes described by Eq.\,(\ref{eq:rho11rho11}) result in coherence changes and phase shifts.

%%%%%%%%%%%%%%%%%%%%%%%

%\subsection{Discussion}\label{ssec:Discussion}

%%%%%%%%%%%%%%%%%%%%%%%%%%%%%%%%%%

\section{Conclusions and outlook}
\label{sec:Conclusion}
The theory of open quantum systems has various applications in many areas of physics. It often makes use of density matrices as its tool of choice for describing open quantum systems and their environments. Tracing out the environmental degrees of freedom leads to reduced density matrices, whose time evolution can be investigated with master equations. However, in many cases, analytically solving a master equation poses an intricate or even impossible task especially if Markovianity is not assumed. 
\\
In the present article we used a previously presented path-integral--based formalism in order to derive, from first principles, a formula that allows us to directly compute reduced density matrices without having to solve a master equation. We only assumed a weak coupling, but allowed for non-Markovian open quantum systems. In addition, the derived formula can be applied to any number of particles in Fock space. We chose to present the formula for density matrix elements in a momentum basis, and for a real scalar field open system interacting with another real scalar field as its environment. However, it is also possible to extend this formalism to other bases like position, and other field species. 
\\
First, we presented the description for density matrices in Fock space and introduced the necessary techniques from non-equilibrium quantum field theory, namely the Feynman-Vernon influence functional, which is based on the Schwinger-Keldysh formalism, and TFD. Next, we stated the well-known fact that, in Schr\"odinger picture TFD, the quantum Liouville equation can be expressed in a Schr\"odinger-like form, which in turn can be solved for a general Hamiltonian. We then applied the LSZ-like reduction formalism that was developed in Ref.\,\cite{Burrage2018} in order to derive the equation that provides a new way of directly computing reduced density matrix elements. Finally, we considered a simple example for an open system and its environment, and applied the formula for computing a few selected density matrix elements for different initial conditions. In this way, we demonstrated the practicability of the presented formalism, and observed open quantum dynamical effects like decoherence. However, we kept the time-dependent divergences, whose curing by renormalization will be discussed in a future article.
\\
Even though we presented the formula in a field theoretical setting, it could well be adapted to non-relativistic quantum physics and find a variety of applications there. In its present form, the formalism discussed here can be applied for providing new insights into phenomena like gravitational or scalar field-induced decoherence, as well as for entirely new investigations in the field theory of open quantum systems. 
%%%%%%%%%%%%%%%%%%%%%%%%%%%%%%%%%%

\begin{acknowledgments}

We thank T.~Colas, P.~Millington and J.~Min\'a\v{r} for useful comments and helpful discussions.
This work was supported by the Austrian Science Fund (FWF): P 34240-N.
\end{acknowledgments}

% =========================================================================================================================================

\bibliography{DMsfFV}

\providecommand{\href}[2]{#2}\begingroup\raggedright\begin{thebibliography}{10}

\bibitem{Breuer2002}
H.-P. Breuer and F.~Petruccione, \emph{The Theory of Open Quantum Systems}.
  Oxford University Press, Oxford, 2002.

\bibitem{Calzetta2008}
E.~A. Calzetta and B.-L. Hu, \emph{Nonequilibrium Quantum Field Theory}.
  Cambridge University Press, Cambridge UK, 2008.

\bibitem{Schlosshauer}
M.~Schlosshauer, \emph{Decoherence and the Quantum-To-Classical Transition}.
  Springer-Verlag Berlin Heidelberg, 2007.

\bibitem{Carmichael}
H.~Carmichael, \emph{An Open Systems Approach to Quantum Optics: Lectures
  Presented at the Universit{\'e} Libre de Bruxelles, October 28 to November 4,
  1991}. Springer Berlin Heidelberg, 1993.

\bibitem{Gardiner2004}
C.~Gardiner and P.~Zoller, \emph{Quantum Noise: A Handbook of Markovian and
  Non-Markovian Quantum Stochastic Methods with Applications to Quantum
  Optics}, Springer Series in Synergetics. Springer, 2004.

\bibitem{Walls2008}
D.~Walls and G.~Milburn, \emph{Quantum Optics}. Springer Berlin Heidelberg,
  2008.

\bibitem{Aolita2015}
L.~Aolita, F.~de~Melo and L.~Davidovich, \emph{Open-system dynamics of
  entanglement:a key issues review},
  \href{https://doi.org/10.1088/0034-4885/78/4/042001}{\emph{Reports on
  Progress in Physics} {\bfseries 78} (2015) 042001}.

\bibitem{Goold2016}
J.~Goold, M.~Huber, A.~Riera, L.~d. Rio and P.~Skrzypczyk, \emph{The role of
  quantum information in thermodynamics—a topical review},
  \href{https://doi.org/10.1088/1751-8113/49/14/143001}{\emph{Journal of
  Physics A: Mathematical and Theoretical} {\bfseries 49} (2016) 143001}.

\bibitem{Werner2016}
A.~Werner, D.~Jaschke, P.~Silvi, M.~Kliesch, T.~Calarco, J.~Eisert et~al.,
  \emph{{Positive Tensor Network Approach for Simulating Open Quantum Many-Body
  Systems}},
  \href{https://doi.org/10.1103/physrevlett.116.237201}{\emph{Physical Review
  Letters} {\bfseries 116} (2016) }.

\bibitem{Huber2020}
J.~Huber, P.~Kirton, S.~Rotter and P.~Rabl, \emph{{Emergence of PT-symmetry
  breaking in open quantum systems}},
  \href{https://doi.org/10.21468/scipostphys.9.4.052}{\emph{SciPost Physics}
  {\bfseries 9} (2020) }.

\bibitem{Koksma2010}
J.~F. Koksma, T.~Prokopec and M.~G. Schmidt, \emph{Decoherence in an
  interacting quantum field theory: The vacuum case},
  \href{https://doi.org/10.1103/PhysRevD.81.065030}{\emph{Phys. Rev. D}
  {\bfseries 81} (2010) 065030}.

\bibitem{Koksma2011}
J.~F. Koksma, T.~Prokopec and M.~G. Schmidt, \emph{Decoherence in an
  interacting quantum field theory: Thermal case},
  \href{https://doi.org/10.1103/PhysRevD.83.085011}{\emph{Phys. Rev. D}
  {\bfseries 83} (2011) 085011}.

\bibitem{Sieberer2016}
L.~M. Sieberer, M.~Buchhold and S.~Diehl, \emph{Keldysh field theory for driven
  open quantum systems}, {\emph{Reports on Progress in Physics} {\bfseries 79}
  (2016) 096001}.

\bibitem{Marino2016}
J.~Marino and S.~Diehl, \emph{Quantum dynamical field theory for nonequilibrium
  phase transitions in driven open systems},
  \href{https://doi.org/10.1103/PhysRevB.94.085150}{\emph{Phys. Rev. B}
  {\bfseries 94} (2016) 085150}.

\bibitem{Baidya2017}
A.~Baidya, C.~Jana, R.~Loganayagam and A.~Rudra, \emph{{Renormalization in open
  quantum field theory. Part I. Scalar field theory}},
  \href{https://doi.org/10.1007/JHEP11(2017)204}{\emph{JHEP} {\bfseries 11}
  (2017) 204} [\href{https://arxiv.org/abs/1704.08335}{{\ttfamily
  1704.08335}}].

\bibitem{Burrage2018}
C.~Burrage, C.~K\"ading, P.~Millington and J.~Min\'a\v{r}, \emph{{Open quantum
  dynamics induced by light scalar fields}},
  \href{https://doi.org/10.1103/PhysRevD.100.076003}{\emph{Phys. Rev. D}
  {\bfseries 100} (2019) 076003}
  [\href{https://arxiv.org/abs/1812.08760}{{\ttfamily 1812.08760}}].

\bibitem{Nagy2020}
S.~Nagy and J.~Polonyi, \emph{{Renormalizing Open Quantum Field Theories}},
  \href{https://doi.org/10.3390/universe8020127}{\emph{Universe} {\bfseries 8}
  (2022) 127} [\href{https://arxiv.org/abs/2012.13811}{{\ttfamily
  2012.13811}}].

\bibitem{Jana2021}
C.~Jana, \emph{{Aspects of open quantum field theory}}, Ph.D. thesis, Tata
  Inst., 2021.

\bibitem{Fogedby2022}
H.~C. Fogedby, \emph{{Field-theoretical approach to open quantum systems and
  the Lindblad equation}},
  \href{https://doi.org/10.1103/PhysRevA.106.022205}{\emph{Phys. Rev. A}
  {\bfseries 106} (2022) 022205}
  [\href{https://arxiv.org/abs/2202.05203}{{\ttfamily 2202.05203}}].

\bibitem{Lombardo1}
F.~Lombardo and F.~D. Mazzitelli, \emph{Coarse graining and decoherence in
  quantum field theory},
  \href{https://doi.org/10.1103/PhysRevD.53.2001}{\emph{Phys. Rev. D}
  {\bfseries 53} (1996) 2001}.

\bibitem{Lombardo2}
F.~C. Lombardo and D.~L. Nacir, \emph{Decoherence during inflation: The
  generation of classical inhomogeneities},
  \href{https://doi.org/10.1103/PhysRevD.72.063506}{\emph{Phys. Rev. D}
  {\bfseries 72} (2005) 063506}.

\bibitem{Lombardo3}
F.~C. Lombardo, \emph{{Influence functional approach to decoherence during
  inflation}},
  \href{https://doi.org/10.1590/S0103-97332005000300005}{\emph{Braz. J. Phys.}
  {\bfseries 35} (2005) 391}
  [\href{https://arxiv.org/abs/gr-qc/0412069}{{\ttfamily gr-qc/0412069}}].

\bibitem{Boyanovsky1}
D.~Boyanovsky, \emph{{Effective field theory during inflation: Reduced density
  matrix and its quantum master equation}},
  \href{https://doi.org/10.1103/PhysRevD.92.023527}{\emph{Phys. Rev.}
  {\bfseries D92} (2015) 023527}
  [\href{https://arxiv.org/abs/1506.07395}{{\ttfamily 1506.07395}}].

\bibitem{Boyanovsky2}
D.~Boyanovsky, \emph{{Effective field theory during inflation. II. Stochastic
  dynamics and power spectrum suppression}},
  \href{https://doi.org/10.1103/PhysRevD.93.043501}{\emph{Phys. Rev.}
  {\bfseries D93} (2016) 043501}
  [\href{https://arxiv.org/abs/1511.06649}{{\ttfamily 1511.06649}}].

\bibitem{Boyanovsky3}
D.~Boyanovsky, \emph{{Fermionic influence on inflationary fluctuations}},
  \href{https://doi.org/10.1103/PhysRevD.93.083507}{\emph{Phys. Rev.}
  {\bfseries D93} (2016) 083507}
  [\href{https://arxiv.org/abs/1602.05609}{{\ttfamily 1602.05609}}].

\bibitem{Boyanovsky4}
D.~Boyanovsky, \emph{{Imprint of entanglement entropy in the power spectrum of
  inflationary fluctuations}},
  \href{https://doi.org/10.1103/PhysRevD.98.023515}{\emph{Phys. Rev.}
  {\bfseries D98} (2018) 023515}
  [\href{https://arxiv.org/abs/1804.07967}{{\ttfamily 1804.07967}}].

\bibitem{Burgess2015}
C.~P. Burgess, R.~Holman, G.~Tasinato and M.~Williams, \emph{{EFT beyond the
  horizon: stochastic inflation and how primordial quantum fluctuations go
  classical}}, \href{https://doi.org/10.1007/JHEP03(2015)090}{\emph{Journal of
  High Energy Physics} {\bfseries 2015} (2015) 90}.

\bibitem{Hollowood}
T.~J. Hollowood and J.~I. McDonald, \emph{Decoherence, discord, and the quantum
  master equation for cosmological perturbations},
  \href{https://doi.org/10.1103/PhysRevD.95.103521}{\emph{Phys. Rev. D}
  {\bfseries 95} (2017) 103521}.

\bibitem{Binder2021}
T.~Binder, K.~Mukaida, B.~Scheihing-Hitschfeld and X.~Yao, \emph{{Non-Abelian
  electric field correlator at NLO for dark matter relic abundance and
  quarkonium transport}},
  \href{https://doi.org/10.1007/JHEP01(2022)137}{\emph{JHEP} {\bfseries 01}
  (2022) 137} [\href{https://arxiv.org/abs/2107.03945}{{\ttfamily
  2107.03945}}].

\bibitem{Yu2008}
H.~W. Yu, J.~Zhang, H.-w. Yu and J.-l. Zhang, \emph{{Understanding Hawking
  radiation in the framework of open quantum systems}},
  \href{https://doi.org/10.1103/PhysRevD.77.029904}{\emph{Phys. Rev. D}
  {\bfseries 77} (2008) 024031}
  [\href{https://arxiv.org/abs/0806.3602}{{\ttfamily 0806.3602}}].

\bibitem{Lombardo2012}
F.~C. Lombardo and G.~J. Turiaci, \emph{{Dynamics of an Acoustic Black Hole as
  an Open Quantum System}},
  \href{https://doi.org/10.1103/PhysRevD.87.084028}{\emph{Phys. Rev. D}
  {\bfseries 87} (2013) 084028}
  [\href{https://arxiv.org/abs/1208.0198}{{\ttfamily 1208.0198}}].

\bibitem{Jana2020}
C.~Jana, R.~Loganayagam and M.~Rangamani, \emph{{Open quantum systems and
  Schwinger-Keldysh holograms}},
  \href{https://doi.org/10.1007/JHEP07(2020)242}{\emph{JHEP} {\bfseries 07}
  (2020) 242} [\href{https://arxiv.org/abs/2004.02888}{{\ttfamily
  2004.02888}}].

\bibitem{Agarwal2020}
K.~Agarwal and N.~Bao, \emph{Toy model for decoherence in the black hole
  information problem},
  \href{https://doi.org/10.1103/PhysRevD.102.086017}{\emph{Phys. Rev. D}
  {\bfseries 102} (2020) 086017}.

\bibitem{Kaplanek2020}
G.~Kaplanek and C.~P. Burgess, \emph{{Qubits on the Horizon: Decoherence and
  Thermalization near Black Holes}},
  \href{https://doi.org/10.1007/JHEP01(2021)098}{\emph{JHEP} {\bfseries 01}
  (2021) 098} [\href{https://arxiv.org/abs/2007.05984}{{\ttfamily
  2007.05984}}].

\bibitem{Burgess2021}
C.~P. Burgess, R.~Holman and G.~Kaplanek, \emph{{Quantum Hotspots: Mean Fields,
  Open EFTs, Nonlocality and Decoherence Near Black Holes}},
  \href{https://arxiv.org/abs/2106.10804}{{\ttfamily 2106.10804}}.

\bibitem{Kaplanek2021}
G.~Kaplanek, C.~P. Burgess and R.~Holman, \emph{{Qubit heating near a
  hotspot}}, \href{https://doi.org/10.1007/JHEP08(2021)132}{\emph{JHEP}
  {\bfseries 08} (2021) 132}
  [\href{https://arxiv.org/abs/2106.10803}{{\ttfamily 2106.10803}}].

\bibitem{Brambilla1}
N.~Brambilla, M.~A. Escobedo, J.~Soto and A.~Vairo, \emph{{Quarkonium
  suppression in heavy-ion collisions: an open quantum system approach}},
  \href{https://doi.org/10.1103/PhysRevD.96.034021}{\emph{Phys. Rev.}
  {\bfseries D96} (2017) 034021}
  [\href{https://arxiv.org/abs/1612.07248}{{\ttfamily 1612.07248}}].

\bibitem{Brambilla2}
N.~Brambilla, M.~A. Escobedo, J.~Soto and A.~Vairo, \emph{{Heavy quarkonium
  suppression in a fireball}},
  \href{https://doi.org/10.1103/PhysRevD.97.074009}{\emph{Phys. Rev.}
  {\bfseries D97} (2018) 074009}
  [\href{https://arxiv.org/abs/1711.04515}{{\ttfamily 1711.04515}}].

\bibitem{Yao2018}
X.~Yao and T.~Mehen, \emph{{Quarkonium in-medium transport equation derived
  from first principles}},
  \href{https://doi.org/10.1103/PhysRevD.99.096028}{\emph{Phys. Rev. D}
  {\bfseries 99} (2019) 096028}
  [\href{https://arxiv.org/abs/1811.07027}{{\ttfamily 1811.07027}}].

\bibitem{Yao2020}
X.~Yao and T.~Mehen, \emph{{Quarkonium Semiclassical Transport in Quark-Gluon
  Plasma: Factorization and Quantum Correction}},
  \href{https://doi.org/10.1007/JHEP02(2021)062}{\emph{JHEP} {\bfseries 02}
  (2021) 062} [\href{https://arxiv.org/abs/2009.02408}{{\ttfamily
  2009.02408}}].

\bibitem{Akamatsu2020}
Y.~Akamatsu, \emph{{Quarkonium in quark\textendash{}gluon plasma: Open quantum
  system approaches re-examined}},
  \href{https://doi.org/10.1016/j.ppnp.2021.103932}{\emph{Prog. Part. Nucl.
  Phys.} {\bfseries 123} (2022) 103932}
  [\href{https://arxiv.org/abs/2009.10559}{{\ttfamily 2009.10559}}].

\bibitem{DeJong2020}
W.~A. De~Jong, M.~Metcalf, J.~Mulligan, M.~P\l{}osko\'n, F.~Ringer and X.~Yao,
  \emph{{Quantum simulation of open quantum systems in heavy-ion collisions}},
  \href{https://doi.org/10.1103/PhysRevD.104.L051501}{\emph{Phys. Rev. D}
  {\bfseries 104} (2021) 051501}
  [\href{https://arxiv.org/abs/2010.03571}{{\ttfamily 2010.03571}}].

\bibitem{Yao2021}
X.~Yao, \emph{{Open quantum systems for quarkonia}},
  \href{https://doi.org/10.1142/S0217751X21300106}{\emph{Int. J. Mod. Phys. A}
  {\bfseries 36} (2021) 2130010}
  [\href{https://arxiv.org/abs/2102.01736}{{\ttfamily 2102.01736}}].

\bibitem{Brambilla2021}
N.~Brambilla, M.~A. Escobedo, M.~Strickland, A.~Vairo, P.~Vander~Griend and
  J.~H. Weber, \emph{{Bottomonium production in heavy-ion collisions using
  quantum trajectories: Differential observables and momentum anisotropy}},
  \href{https://doi.org/10.1103/PhysRevD.104.094049}{\emph{Phys. Rev. D}
  {\bfseries 104} (2021) 094049}
  [\href{https://arxiv.org/abs/2107.06222}{{\ttfamily 2107.06222}}].

\bibitem{Griend2021}
P.~V. Griend, \emph{{Bottomonium observables in an open quantum system using
  the quantum trajectories method}},
  \href{https://doi.org/10.1051/epjconf/202225805005}{\emph{EPJ Web Conf.}
  {\bfseries 258} (2022) 05005}
  [\href{https://arxiv.org/abs/2111.13520}{{\ttfamily 2111.13520}}].

\bibitem{Yao2022}
X.~Yao, \emph{{Quarkonium Suppression in the Open Quantum System Approach}},
  in \emph{{19th International Conference on Hadron Spectroscopy and
  Structure}}, 1, 2022, \href{https://arxiv.org/abs/2201.07702}{{\ttfamily
  2201.07702}}.

\bibitem{Blencowe}
M.~P. Blencowe, \emph{{Effective Field Theory Approach to Gravitationally
  Induced Decoherence}},
  \href{https://doi.org/10.1103/PhysRevLett.111.021302}{\emph{Phys. Rev. Lett.}
  {\bfseries 111} (2013) 021302}
  [\href{https://arxiv.org/abs/1211.4751}{{\ttfamily 1211.4751}}].

\bibitem{Anastopoulos2013}
C.~Anastopoulos and B.~L. Hu, \emph{{A Master Equation for Gravitational
  Decoherence: Probing the Textures of Spacetime}},
  \href{https://doi.org/10.1088/0264-9381/30/16/165007}{\emph{Class. Quant.
  Grav.} {\bfseries 30} (2013) 165007}
  [\href{https://arxiv.org/abs/1305.5231}{{\ttfamily 1305.5231}}].

\bibitem{Oniga2015}
T.~Oniga and C.~H.~T. Wang, \emph{{Quantum gravitational decoherence of light
  and matter}}, \href{https://doi.org/10.1103/PhysRevD.93.044027}{\emph{Phys.
  Rev. D} {\bfseries 93} (2016) 044027}
  [\href{https://arxiv.org/abs/1511.06678}{{\ttfamily 1511.06678}}].

\bibitem{Minar2016}
J.~Min\'a\v{r}, P.~Sekatski and N.~Sangouard, \emph{Bounding
  quantum-gravity-inspired decoherence using atom interferometry},
  \href{https://doi.org/10.1103/PhysRevA.94.062111}{\emph{Phys. Rev. A}
  {\bfseries 94} (2016) 062111}.

\bibitem{Minar2016_2}
J.~Min\'a\v{r}, P.~Sekatski, R.~Stevenson and N.~Sangouard, \emph{Testing
  unconventional decoherence models with atoms in optical lattices},  2016.

\bibitem{Bassi2017}
A.~Bassi, A.~Gro\ss{}ardt and H.~Ulbricht, \emph{{Gravitational Decoherence}},
  \href{https://doi.org/10.1088/1361-6382/aa864f}{\emph{Class. Quant. Grav.}
  {\bfseries 34} (2017) 193002}
  [\href{https://arxiv.org/abs/1706.05677}{{\ttfamily 1706.05677}}].

\bibitem{Asprea2019}
L.~Asprea, G.~Gasbarri and A.~Bassi, \emph{{Gravitational decoherence: A
  general nonrelativistic model}},
  \href{https://doi.org/10.1103/PhysRevD.103.104041}{\emph{Phys. Rev. D}
  {\bfseries 103} (2021) 104041}
  [\href{https://arxiv.org/abs/1905.01121}{{\ttfamily 1905.01121}}].

\bibitem{Asprea2020}
L.~Asprea, A.~Bassi, H.~Ulbricht and G.~Gasbarri, \emph{Gravitational
  decoherence and the possibility of its interferometric detection},
  \href{https://doi.org/10.1103/PhysRevLett.126.200403}{\emph{Phys. Rev. Lett.}
  {\bfseries 126} (2021) 200403}.

\bibitem{Asprea2021}
L.~Asprea and G.~Gasbarri, \emph{Gravitational decoherence: A nonrelativistic
  spin 1/2 fermionic model},
  \href{https://doi.org/10.1103/PhysRevD.104.024043}{\emph{Phys. Rev. D}
  {\bfseries 104} (2021) 024043}.

\bibitem{Lagouvardos2021}
M.~Lagouvardos and C.~Anastopoulos, \emph{Gravitational decoherence of
  photons}, \href{https://doi.org/10.1088/1361-6382/abf2f3}{\emph{Classical and
  Quantum Gravity} {\bfseries 38} (2021) 115012}.

\bibitem{Anastopoulos2021}
C.~Anastopoulos and B.-L. Hu, \emph{{Gravitational decoherence: A thematic
  overview}}, \href{https://doi.org/10.1116/5.0077536}{\emph{AVS Quantum Sci.}
  {\bfseries 4} (2022) 015602}
  [\href{https://arxiv.org/abs/2111.02462}{{\ttfamily 2111.02462}}].

\bibitem{Pikovski2013}
I.~Pikovski, M.~Zych, F.~Costa and {\v{C}}.~Brukner, \emph{{Universal
  decoherence due to gravitational time dilation}},
  \href{https://doi.org/10.1038/nphys3366}{\emph{Nature Phys.} {\bfseries 11}
  (2015) 668} [\href{https://arxiv.org/abs/1311.1095}{{\ttfamily 1311.1095}}].

\bibitem{Pikovski2017}
I.~Pikovski, M.~Zych, F.~Costa and {\v{C}}.~Brukner, \emph{Time dilation in
  quantum systems and decoherence},
  \href{https://doi.org/10.1088/1367-2630/aa5d92}{\emph{New Journal of Physics}
  {\bfseries 19} (2017) 025011}.

\bibitem{Burrage2019}
C.~Burrage, C.~K\"ading, P.~Millington and J.~Min\'a\v{r}, \emph{{Influence
  functionals, decoherence and conformally coupled scalars}},
  \href{https://doi.org/10.1088/1742-6596/1275/1/012041}{\emph{J. Phys. Conf.
  Ser.} {\bfseries 1275} (2019) 012041}
  [\href{https://arxiv.org/abs/1902.09607}{{\ttfamily 1902.09607}}].

\bibitem{Kading2019}
C.~K\"ading, \emph{{Astro- and Quantum Physical Tests of Screened Scalar
  Fields}}, Ph.D. thesis, University of Nottingham, Nottingham NG7 2RD, UK, 10,
  2019.
\newblock \href{https://arxiv.org/abs/1910.05738}{{\ttfamily 1910.05738}}.

\bibitem{Lehmann1954}
H.~Lehmann, K.~Symanzik and W.~Zimmermann, \emph{{On the formulation of
  quantized field theories}},
  \href{https://doi.org/10.1007/BF02731765}{\emph{Nuovo Cim.} {\bfseries 1}
  (1955) 205}.

\bibitem{Takahasi:1974zn}
Y.~Takahasi and H.~Umezawa, \emph{{Thermo field dynamics}}, {\emph{Collect.
  Phenom.} {\bfseries 2} (1975) 55}.

\bibitem{Arimitsu:1985ez}
T.~Arimitsu and H.~Umezawa, \emph{{A General Formulation of Nonequilibrium
  Thermo Field Dynamics}},
  \href{https://doi.org/10.1143/PTP.74.429}{\emph{Prog. Theor. Phys.}
  {\bfseries 74} (1985) 429}.

\bibitem{Arimitsu:1985xm}
T.~Arimitsu and H.~Umezawa, \emph{{Non-Equilibrium Thermo Field Dynamics}},
  \href{https://doi.org/10.1143/PTP.77.32}{\emph{Prog. Theor. Phys.} {\bfseries
  77} (1987) 32}.

\bibitem{Khanna}
F.~C. Khanna, A.~P.~C. Malbouisson, J.~M.~C. Malbouisson and A.~E. Santana,
  \emph{Thermal Quantum Field Theory: Algebraic Aspects and Applications}.
  World Scientific, Singapore, 2009.

\bibitem{Schwinger}
J.~S. Schwinger, \emph{{Brownian Motion of a Quantum Oscillator}},
  \href{https://doi.org/10.1063/1.1703727}{\emph{J. Math. Phys.} {\bfseries 2}
  (1961) 407}.

\bibitem{Keldysh}
L.~V. Keldysh, \emph{{Diagram technique for nonequilibrium processes}},
  {\emph{Zh. Eksp. Teor. Fiz.} {\bfseries 47} (1964) 1515}.

\bibitem{Feynman}
R.~P. Feynman and F.~L. Vernon, \emph{The theory of a general quantum system
  interacting with a linear dissipative system}, {\emph{Annals of physics}
  {\bfseries 24} (1963) 118}.

\bibitem{LeBellac}
M.~Le~Bellac, \emph{Thermal field theory}. Cambridge University Press, 1996.

\bibitem{Loubenets2020}
E.~R. Loubenets and C.~Käding, \emph{{Specifying the Unitary Evolution of a
  Qudit for a General Nonstationary Hamiltonian via the Generalized Gell-Mann
  Representation}}, \href{https://doi.org/10.3390/e22050521}{\emph{Entropy}
  {\bfseries 22} (2020) }.

\bibitem{Millington:2012pf}
P.~Millington and A.~Pilaftsis, \emph{{Perturbative nonequilibrium thermal
  field theory}}, \href{https://doi.org/10.1103/PhysRevD.88.085009}{\emph{Phys.
  Rev.} {\bfseries D88} (2013) 085009}
  [\href{https://arxiv.org/abs/1211.3152}{{\ttfamily 1211.3152}}].

\bibitem{Millington:2013isa}
P.~Millington and A.~Pilaftsis, \emph{{Perturbative Non-Equilibrium Thermal
  Field Theory to all Orders in Gradient Expansion}},
  \href{https://doi.org/10.1016/j.physletb.2013.05.044}{\emph{Phys. Lett.}
  {\bfseries B724} (2013) 56}
  [\href{https://arxiv.org/abs/1304.7249}{{\ttfamily 1304.7249}}].

\bibitem{Wick}
G.~C. Wick, \emph{The evaluation of the collision matrix},
  \href{https://doi.org/10.1103/PhysRev.80.268}{\emph{Phys. Rev.} {\bfseries
  80} (1950) 268}.

\bibitem{Kading2022}
C.~K\"ading, P.~Millington and M.~Pitschmann, \emph{to be published},
  \href{https://arxiv.org/abs/2xxx.xxxxx}{{\ttfamily 2xxx.xxxxx}}.

\end{thebibliography}\endgroup
\bibliographystyle{JHEP}

\end{document}